\documentclass[fleqn,usenatbib,useAMS]{mnras}

\usepackage{graphicx}	
\usepackage{subcaption}
\usepackage{amsmath}	
\usepackage{amssymb}	
\usepackage{multicol}        
\usepackage{bm}		
\usepackage{pdflscape}	
\usepackage{enumitem}
\usepackage{animate}

\usepackage[T1]{fontenc}
\usepackage{ae,aecompl}

\usepackage{newtxtext,newtxmath}

\title[Variance of galaxy spectra]{What drives the variance of galaxy spectra?}

\author[Sharbaf, Ferreras \& Lahav]{Zahra Sharbaf$^{1,2}$\thanks{E-mail: \href{mailto:zsharbaf@iac.es}{zsharbaf@iac.es}}, Ignacio Ferreras$^{1,2,3}$\thanks{E-mail: \href{mailto:i.ferreras@ucl.ac.uk}{i.ferreras@ucl.ac.uk}}, 
  Ofer Lahav$^3$\\
$^{1}$Instituto de Astrofìsica de Canarias, C/ Vìa La ctea s/n, La Laguna, E-38200 La Laguna, Tenerife, Spain
\\
$^{2}$Departamento de Astrofsica, Universidad de La Laguna, E-38205 La Laguna, Tenerife, Spain
\\
$^{3}$Department of Physics and Astronomy, University College London, Gower Street, London WC1E 6BT, UK\\
}

\date{Published in MNRAS, Accepted 2023 August 31. Received 2023 August 31; in original form 2023 May 9}
\pubyear{2023}

\begin{document}
\label{firstpage}
\pagerange{\pageref{firstpage}--\pageref{lastpage}}
\maketitle

\begin{abstract}
We present a study aimed at understanding the physical phenomena
underlying the formation and evolution of galaxies following a
data-driven analysis of spectroscopic data based on the variance in a
carefully selected sample. We apply Principal Component Analysis (PCA)
independently to three subsets of continuum-subtracted optical
spectra, segregated into their nebular emission activity as quiescent,
star-forming, and Active Galactic Nuclei (AGN). We emphasize that the
variance of the input data in this work only relates to the absorption
lines in the photospheres of the stellar populations. The sample is
taken from the Sloan Digital Sky Survey (SDSS) in the stellar velocity
dispersion range 100--150\,km/s, to minimise the ``blurring'' effect
of the stellar motion. We restrict the analysis to the first three
principal components (PCs), and find that PCA segregates the three
types with the highest variance mapping SSP-equivalent age, along with
an inextricable degeneracy with metallicity, even when all three PCs
are included. Spectral fitting shows that stellar age dominates PC1,
whereas PC2 and PC3 have a mixed dependence of age and metallicity.
The trends support -- independently of any model fitting -- the
hypothesis of an evolutionary sequence from star-formation to AGN to
quiescence. As a further test of the consistency of the analysis, we
apply the same methodology in different spectral windows, finding
similar trends, but the variance is maximal in the blue wavelength
range, roughly around the 4000\AA\ break.
\end{abstract}


\section{Introduction}
\label{Sec:Intro}

The formation history of galaxies represents a record of the various physical
processes that lead to the stellar and gaseous components that we 
observe at the telescope.  In this framework, galaxy spectra
encode the kinematics, age and chemical composition of the underlying stellar
populations, and thus provide the best available source of information
in extragalactic astrophysics. However, retrieving this information is
fraught with problems due to the large entanglement of the individual
components, preventing us from producing detailed constraints, other than
overall estimates such as average ages and metallicities. 
Over the now long history of stellar
population modelling, large advances have been
produced
\citep[well documented in many reviews, to name a few:][]{Renzini:06,Walcher:11,Conroy:13},
and population synthesis models have become commonplace in the analyses of
galaxy data
\citep[see, e.g.][]{Bruzual:93,W:94,BC:03,Fioc:97,Maraston:05,Maraston:11,Vazdekis:10,Conroy:10}.
The building block of these models is the simple stellar population (SSP), an
ensemble of stars that are all formed at the same time, with the same
chemical composition, following a well-defined distribution of stellar
mass (the so-called initial mass function).
Any formation history can be described by a superposition of
SSPs, and the resulting photometric and spectroscopic output can be compared
with the observations. Three techniques are typically 
applied to this problem: 1) Model fitting: the spectroscopic
information is compared with population synthesis models, whose
parameters can be treated in a Bayesian framework, producing estimates
of average stellar age, metallicity, abundance ratios, quenching
timescales, etc. \citep[see, e.g.,][]{Trager:00,Gallazzi:05,FLB:14}; 2) Machine Learning methods have recently come
to the fore, both in a supervised and unsupervised way, with the goal
of extracting the kind of information that traditional methods cannot
produce \citep[e.g.][]{Ball:04,Lovell:19,Portillo:20,Melchior:22,MHC:23}; 3) Multivariate analysis methods
can be applied to the spectroscopic data, reducing the problem to blind source separation,
making use of data-driven diagnostics such as covariance, clustering,
or statistical independence, to overcome the
entanglement \citep[e.g.][]{Madgwick:03,Kaban:05,Lu:06,Ferreras:06,Nolan:07,Rogers:10,Wild:14}.

Comparisons with population synthesis models have produced fundamental
constraints in galaxy properties, such as the correlation between age
or metallicity and stellar mass \citep[e.g.][]{Gallazzi:05}; the
increasing [Mg/Fe] in massive galaxies as a signature of rapid
formation \citep[e.g.][]{Trager:00, Thomas:05, IGDR:11}; the differing
role of environment in central and satellite galaxies
\citep[e.g.][]{Pasquali:10,Peng:12,FLB:14}; or the presence of small
episodes of star formation in early-type galaxies
\citep[e.g.][]{FS:00,Kaviraj:07,Salvador:20}.  However, population
synthesis methods still produce general results that cannot be taken
beyond the lowest order moments of the distributions in age and
chemical composition. For instance, quenching timescales in green
valley galaxies determined by direct modelling are still rather 
vague \citep[e.g.,][]{Schawinski:14,Nogueira:18,Angthopo:20}
and all too often
comparable with the typical evolutionary timescales over which most
of the stellar populations vary substantially in the sense of
information content \citep{InfoPop}. Different spectral
fitting codes can produce acceptable solutions with very different
formation histories \citep{Ge:18,Ge:19}.

This paper belongs to probe number 3 in the above list, adopting
covariance as the main descriptor of information in galaxy spectra,
and making use of Principal Component Analysis (PCA) to understand the
main source of variations in the observational data from the Sloan
Digital Sky Survey. Furthermore, instead of applying our methodology
to synthetic data, we focus on the information content of the
observational spectra, in a fully data-driven way. We only use models
(e.g. spectral fits to simple stellar populations) a posteriori to
give physical meaning to the components statistically derived from the
observations. This approach inescapably avoids the prior definition of
parameters (e.g. age, metallicity), but, rather, aims at looking for
the ``signal'' present in a carefullty selected sample of spectra.
This project is thus complementary to standard techniques, where 
the goal is to explore at a more fundamental level the limitation
of spectral analysis in galaxies.

An important specificity of our work is that we independently apply
PCA to three subsets of spectra, classified by their nebular emission
into three groups, namely star-forming, AGN, and quiescent galaxies;
as explained in sections~\ref{Sec:Dec} and \ref{Sec:Disc}. While the
input spectra originate from the same dataset, a separate analysis allows
us to explore the individual properties of these three groups, and to assess
the way variance is distributed  in the three classes.
This new approach to PCA builds upon earlier works that also targeted the 
spectral energy distribution of galaxies
\citep[e.g][]{Yip:04,Ferreras:06,Rogers:07,Portillo:20,Tous:20}.
An additional novelty to our analysis is that the
continuum is removed -- to focus on the signal from the absorption lines, and
spectral windows affected by strong  nebular emission lines are
removed from the analysis. Moreover, conservative culling is applied to suppress
discordant data before determining the covariance matrix in order to fully concentrate 
on variations caused by the stellar populations (see sections \ref{Sec:Sample}
and \ref{Sec:PCA}).

The structure of the paper is as follows: the sample is presented in \S\,\ref{Sec:Sample},
followed by an overview of the methodology in \S\,\ref{Sec:PCA}. The spectral data
are decomposed into principal components in \S\,\ref{Sec:Dec}, and the projections are
explored in \S\,\ref{Sec:Proj}, along with models of population synthesis. Finally,
we discuss the results and present our conclusions in \S\,\ref{Sec:Disc}.

\section{Sample} 
\label{Sec:Sample}

The spectroscopic sample adopted in this paper is taken from the Sloan
Digital Sky Survey \citep[SDSS, ][]{SDSS}. We use the Legacy
dataset comprising single fibre spectroscopy at resolution
R$\sim$2,000 \citep{Smee:13}, from Data Release 16
\citep{DR16}. Systematic effects will be easily picked up by
any data-driven method based on variance, so that samples from homogeneous, well-calibrated
surveys are essential to avoid spurious results.
Moreover, general samples of galaxy spectra will feature a strong
variance regarding the well-known correlation of the stellar
population properties with stellar mass or velocity dispersion
\citep{Bernardi:03,Gallazzi:05,Gallazzi:14,Ferreras:19}. Therefore, in
order to select a more focused sample, we restrict the stellar
velocity dispersion in the range $\sigma\in[100,150]$\,km/s. We also
restrict the redshift to $z\in[0.05,0.1]$ to minimise aperture
effects, and impose a threshold on the signal to noise ratio ($>$15
per $\Delta\log(\lambda/$\AA$)=10^{-4}$ pixel in the SDSS-r band ) to
reduce the contribution from noise to the variance. The final sample
comprises 68,794 spectra.

The spectra were dereddened and deredshifted using a linear
interpolation algorithm with redshift and foreground dust estimates
supplied by SDSS.  Finally, the SEDs were normalized to the same
average flux across the 6000–6500\AA\ rest-frame wavelength range.  In
this work we try to minimise as much as possible the potential
systematics. Therefore, at the cost of discarding part of the
information content of spectra, we remove the continuum.  This will
allow us to eliminate systematics caused by dust reddening as well as
by flux calibration residuals (over similar wavelength scales). We
measure the pseudo-continuum with the high percentile method laid out
in \citet{BMC:10}, liberally defined as the boosted median continuum
(BMC). We adopt a 100\AA\ running window and a 95\% level, an optimal
choice for SDSS spectra \citep[see also][]{Hawkins:14}. In addition,
once PCA is applied to the data, we explore the standard deviation
of the distribution as a function of wavelength to pinpoint problematic
data, resulting in a reduced size of the sample, as discussed
in detail in the next section.
As we explore stellar populations through the variance of the input
data, we focus our analysis on the spectral interval 3800-4200\AA --
which is the region that keeps most of the information content 
\citep[see, e.g.,][]{InfoPop}. However, as reference, we also consider the 
5000-5400\AA\ window -- that includes prominent Mg and Fe spectral features.

\section{Principal Component Analysis (PCA)}
\label{Sec:PCA}

PCA, closely related to the Karhunen–Lo\`{e}ve theorem \citep{Karhunen:47}, is a widely
adopted multivariate method to explore the variance of large datasets and
can be used to classify, compress, denoise data, or to produce a lower
dimensional latent space that can be further applied to more sophisticated
techniques. PCA rearranges the input data -- each defined by N numbers,
here the fluxes within a range of N wavelength intervals --  into a ranked set of
variables, the principal components,  by performing rotations in 
N-dimensional parameter space. These components are  defined as those that diagonalise the
covariance matrix of the data. Hence, by definition, the principal components
are decorrelated -- note that, in general, this does not mean statistical
independence. A number of papers have been devoted to explore PCA in
galaxy spectra \citep[to name a few:][]{Yip:04,Ferreras:06,Rogers:07,Portillo:20,Tous:20}.

In a nutshell, PCA reduces to the diagonalisation of the covariance
matrix, where the eigenvalues represent the individual contribution to
the variance of the eigenvectors (also known as principal
components). These are ranked in decreasing order of
variance, typically given as the fractional contribution, producing
the so-called scree plot (fractional variance vs rank), that gives a
graphical representation of how higher order components contribute
progressively less towards the total variance. Projecting the spectra
on to the eigenvectors produces the ``coordinates'' in latent space,
and the first few components are expected to keep most of the information, in
the sense of variance.  Although PCA is mostly used for classification purposes 
\citep[e.g.,][]{folkes1996, Madgwick:03, Nersesian:21}, in this
study we target instead how the latent space encodes differences in
the underlying stellar populations of different groups of galaxies.
More information about this approach to PCA in galaxy spectra can be
found in \citet{Ferreras:06} and \citet{Rogers:07,Rogers:10}.  The
samples in those papers focused on quiescent, early-type galaxies,
whereas the present work does not introduce such a restriction,
including star-forming and AGN galaxies in the analysis.

\begin{figure*}
\includegraphics[width=80mm]{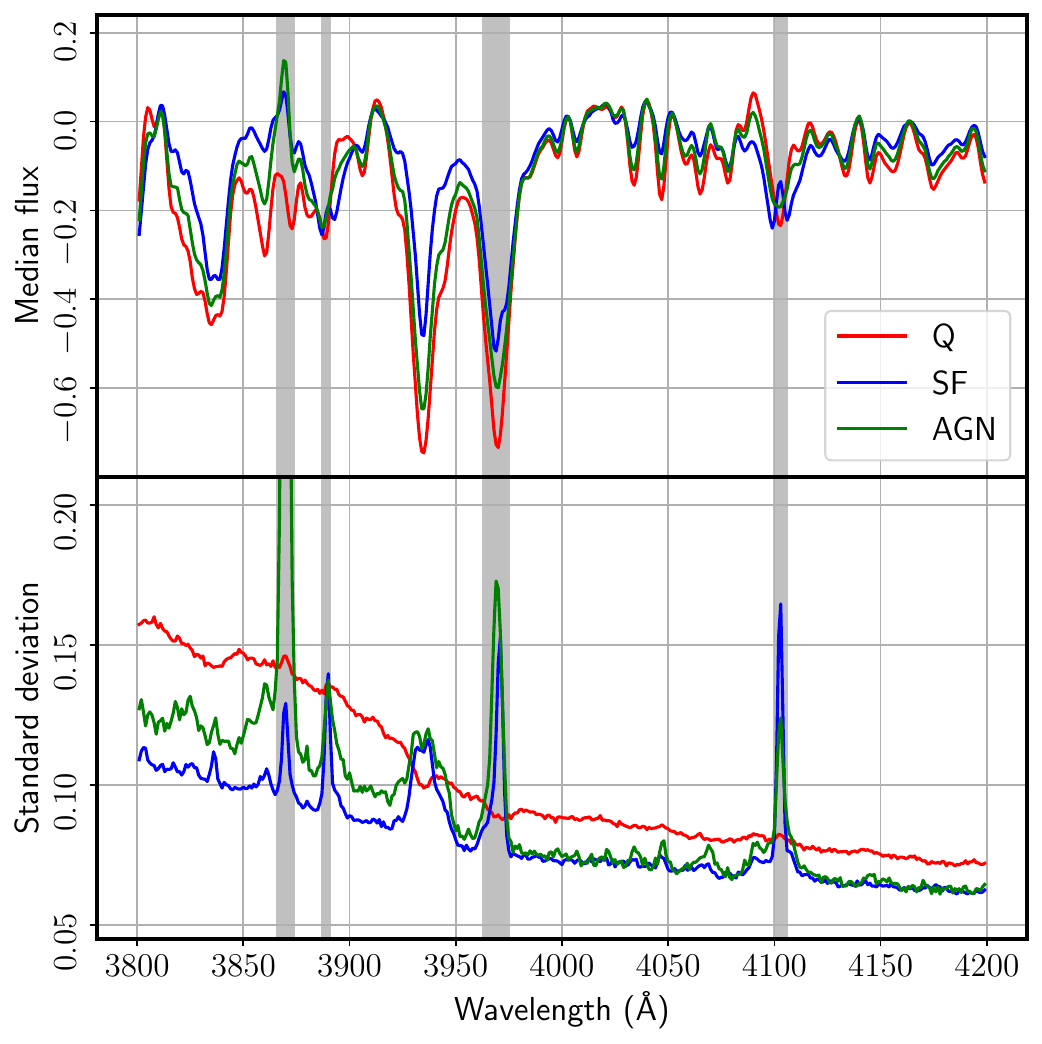}
\includegraphics[width=80mm]{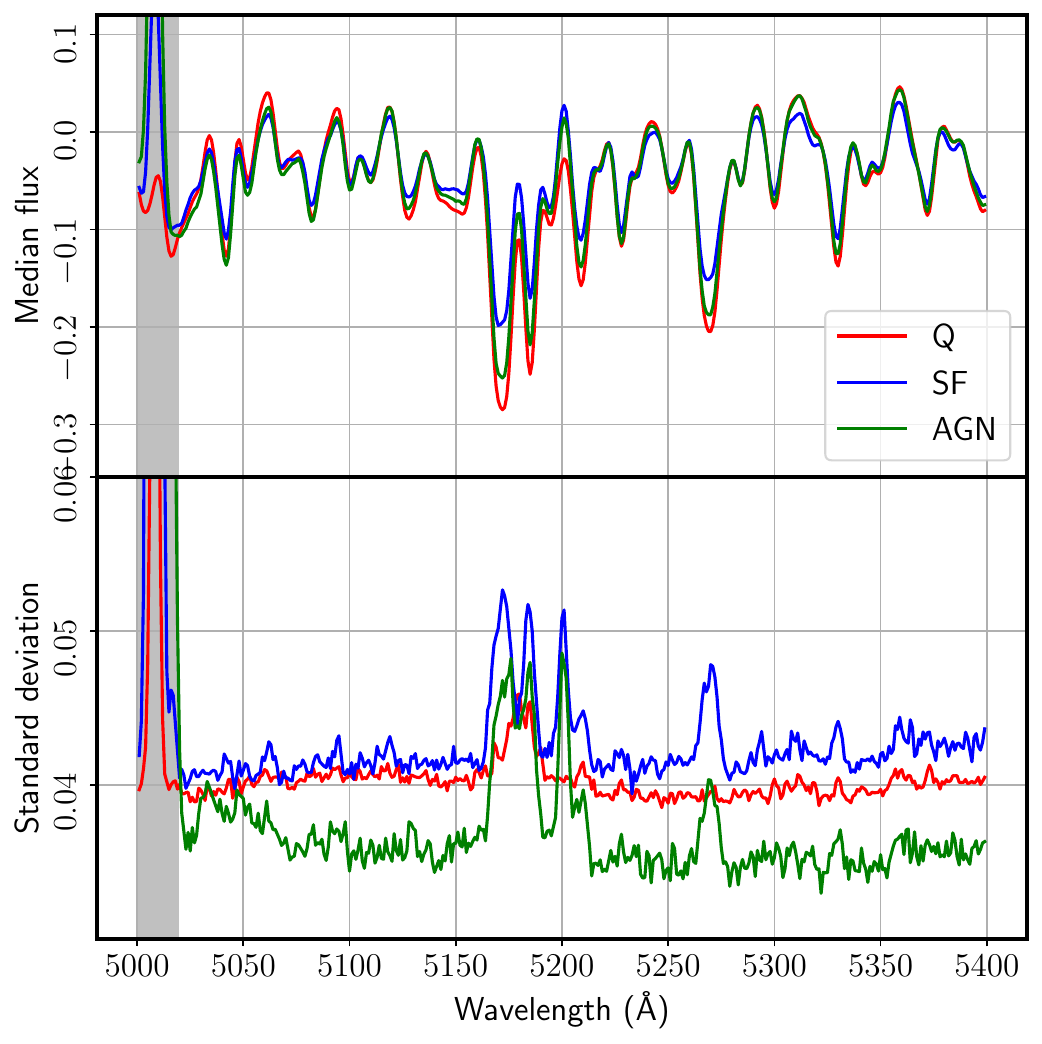}
\caption{Median (top) and standard deviation (bottom) of the SDSS
  continuum-subtracted spectra
  used in this paper (including a 4$\sigma$ cull of the input data),
  segregated with respect to nebular activity
  into quiescent (Q), star-forming (SF) and AGN. The grey shaded regions
represent the intervals removed from the analysis.}
\label{fig:SigPlot}
\end{figure*}

One can interpret a spectrum as a vector in a high dimensional space
spanned by a set of wavelengths ($\lambda_i$, $i=\{1,\cdots,N\}$).
Therefore, the fluxes corresponding to the $k$-th galaxy in a sample
(${\Phi_k(\lambda_i)}$, $k=\{1,\cdots,M\}$) are the coordinates in the
space spanned by unit vectors $\hat{\lambda}_i$.  Our aim is to
derive a lower-dimensional set of vectors that are representative of a large sample 
of galaxy spectra, so that the reduced set of components encapsulates
most of the variance in the ensemble.  The fundamental element in this method is the
difference between the flux of a spectrum, say for the $k$th galaxy,
and the mean at a given wavelength ($\lambda_i$):
\begin{equation}
\Delta_k(\lambda_i) \equiv 
\Bigl(\Phi_k(\lambda_i) - \langle\Phi_s(\lambda_i)\rangle_s\Bigr),
\end{equation}
where $\langle\cdots\rangle_s$ represents averaging the quantity inside the brackets,
with respect to the sample of ($s$=$\{1,\cdots,M\}$) galaxy spectra. 
The covariance is defined as follows:
\begin{equation}
  {\cal C}_{ij}\equiv \biggl\langle \Delta_k(\lambda_i) \Delta_k(\lambda_j)
  \biggl\rangle_k.
\end{equation}
This is an $N\times N$ symmetric, positive-definite matrix. 
We note that the covariance is sensitive to the presence of outliers
-- an equivalent case is least squares fitting a set of data points to
some model function. To produce a more robust covariance,
we apply a conservative culling process, where spectra are removed
if the following condition applies:
\begin{equation}
|\Delta_k(\lambda_i)| > 4\sigma(\lambda_i),
\label{eq:4sigma}
\end{equation}
where $\sigma(\lambda_i)$ is the standard deviation of the distribution
of fluxes at a given wavelength. Note this condition is enforced at all
wavelengths in the spectral region of interest. Therefore, the adopted culling 
process is rather aggressive, but ensures that our PCA results are safely
driven by the underlying variance of the data and not by the presence
of outliers. In order to diagonalise the covariance matrix, we need to solve
the eigenvalue problem, that involves finding a matrix ${\cal U}$,
such that:
\begin{equation}\label{eq2}
{\cal U}^{-1}{\cal C}{\cal U}=\Pi
\end{equation}
where $\Pi$ is a diagonal matrix containing the eigenvalues of ${\cal C}$, 
and {${\cal U}$} contains the eigenspectra (or principal components), e.g.:
\[
\hat{\bm{e}}_k=\sum_j{\cal U}_{j,k}\hat{\lambda}_j, {\rm such\ that}\quad {\cal C}\hat{\bm{e}}_k=\pi_k\hat{\bm{e}}_k,
\]
with $\pi_k$ representing the eigenvalues (i.e. the variance associated
to the $k$th principal component).
Now we can project the observed spectra
on the $\{\hat{e}_k\}$ vectors to obtain the ``coordinates'', eigencoefficients, or -- overloading notation -- ``principal components''.
Therefore, we now have two alternative representations of a galaxy
spectrum: the traditional one, given by the fluxes,
$\{\Phi(\lambda_i)\}$, or the principal component projections,
$\{{\rm PC}_k\}$. The advantage of the latter is that we can restrict the
description to the first few components and still keep most of the
original variance of the dataset. Note that, similarly to least
squares fitting, PCA is overly sensitive to outliers -- the former
indeed minimises the variance between the data and some fitting
function. Therefore, it is very important to ensure the input data are
free from systematics, hence the careful definition of the sample
presented in the previous section. For instance, galaxy spectra will
be affected by emission lines from the diffuse gas, and spectra with
very intense emission lines (as in star-bursting galaxies or strong
AGN), will disproportionately affect the first few principal
components, biasing the analysis.  Our goal is to assess differences
concerning the stellar populations, leading us to avoid a few spectral
windows with prominent emission lines that may dominate the
variance. Rather than masking the regions, we completely remove these
spectral intervals when computing the covariance matrix. 

\begin{figure*}
\includegraphics[width=80mm]{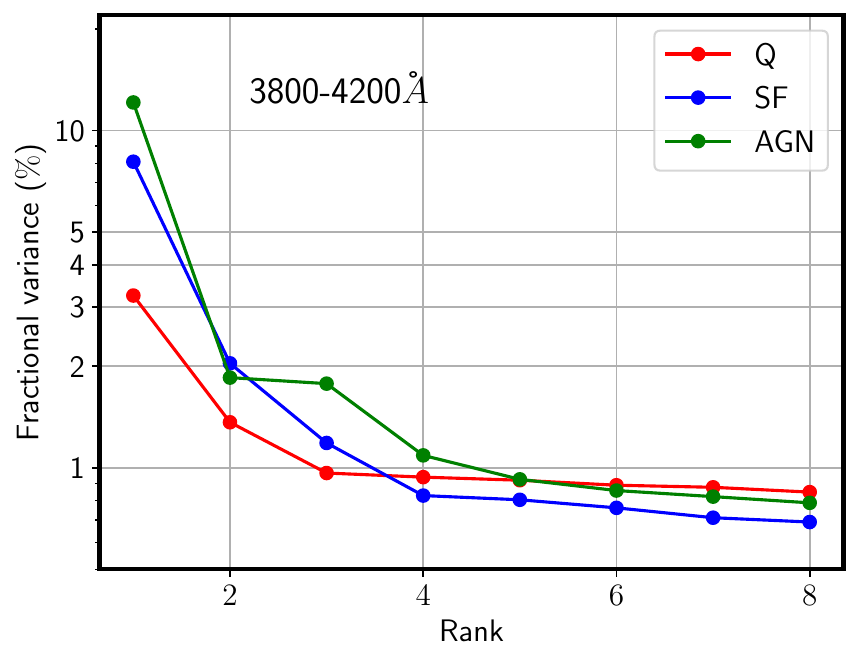}
\includegraphics[width=80mm]{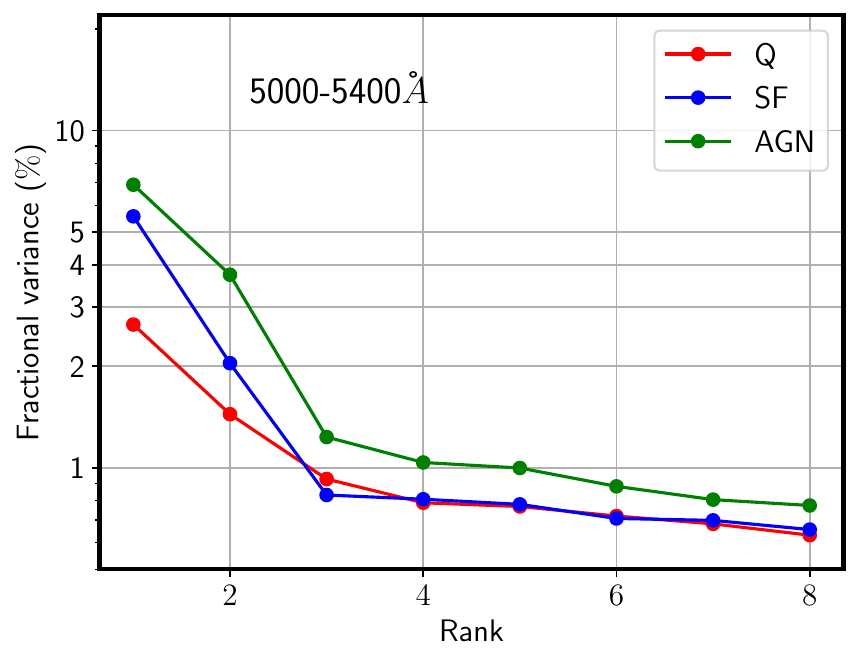}
\caption{Scree plots showing the first eight eigenvalues of the principal component
  analysis decomposition performed in the blue (left) and red (right) spectral intervals.
  The sample is split regarding nebular emission activity into quiescent (red, Q),
star-forming (blue, SF) and AGN (green), as labelled.}
\label{fig:scree}
\end{figure*}

\section{Decomposition of SDSS spectra into principal components}
\label{Sec:Dec}

In order to explore the stellar populations of 
different groups of galaxies through the variance of the input
data, we focus our study on two spectral intervals. 
These intervals keep most of the variance from galaxy to
galaxy \citep[see, e.g.,][]{InfoPop}.
The dominant region is roughly centered on the 4000\AA\ break, between 
3800 and 4200 {\AA}. This region features a large number of
absorption lines that strongly depend on the age and chemical composition
of the underlying populations \citep[see, e.g.,][]{BC:03}. The second interval is defined between
5000 and 5400\AA\, and mainly focuses
on the Mg--Fe complex, a region often used to constrain chemical abundance.
Throughout this paper we will refer to these two spectral regions as the ``blue''
and ``red'' interval, respectively.
Fig.~\ref{fig:SigPlot} shows the median (top) and standard deviation
(bottom) of the continuum subtracted spectra, segregated into
quiescent (Q), star-forming (SF) and AGN, as labelled, in the blue and
red intervals.  Note the  contribution to the variance of the
most prominent emission lines: 3869\AA ([Ne{\sc iii}]); 3889\AA
(H$\zeta$); 3969\AA (H$\epsilon$) and 4100\AA (H$\delta$), in the
blue interval, and 5007\AA\ ([O{\sc iii}]) in the red interval, 
shown as grey shaded regions. Since these lines originate in the
diffuse gas, we remove the spectral windows from the covariance matrix. 
The reason for removing these regions can be seen in the standard
deviation plot of Figure~\ref{fig:SigPlot}, 
the emission lines feature strong variations among galaxies.
Such a signal would be picked up by PCA as a source of information
or variance and influence our analysis. 
The size of the avoided regions is chosen so as to fully eliminate the contribution
of the emission lines, as shown in Fig.~\ref{fig:SigPlot}. 
We emphasize that our aim is to avoid any source of variance unrelated to the
stellar content, not to achieve a level of completeness. 
The remaining spectral features contributing to the variance originate in
the stellar atmospheres and show up as absorption lines.

While the rest-frame NUV region contributes a rich selection of population sensitive absorption 
lines \citep[e.g.,][]{AV:16}, we do not consider it, as the S/N of SDSS spectra
drops  blueward of $\lambda\lesssim$3800\AA, heavily affecting
any variance analysis.
At the red end, the variance of the absorption lines beyond
5400\AA\ is substantially lower. However, we show further below a comparison of these
two spectral intervals with an additional third one, 5800-6200\AA, to illustrate the 
consistency of our analysis. We do not extend the analysis beyond 
6200\AA, so as not to be affected by
the strong lines H$\alpha$ and [N{\sc ii}] emitted by the 
diffuse gas. Note that airglow emission lines might
introduce prominent residuals in individual spectra, however our working sample
covers quite homogeneously a wide range of redshifts, thus minimising the chance of sky emission
lines affecting the combined analysis (see Appendix~\ref{Sec:ApSky}). It is
worth noting that PCA can be alternatively applied in the observer
frame, in order to remove the contribution from airglow \citep{WH:05}.

The SDSS spectra has been classified with respect to nebular emission into
star forming, AGN and quiescent galaxies, making use of the standard BPT line ratio
diagnostics \citep{BPT}. While our analysis strives to eliminate the
contribution from dust and gas, the BPT classification has a further hypothetical interpretation
as an evolutionary transition from star formation towards quiescence
\citep[see, e.g.][]{Schawinski:07, Angthopo:19}.
We bin the galaxy spectra according to this criterion to focus
our analysis on three separate stages of galaxy evolution. Hence, we 
define three independent covariance matrices corresponding to spectra classified as
star forming (SF), AGN, and quiescent (Q). We emphasize that PCA is a powerful
data-driven technique that efficiently detects the different contributions of
variance. However, it is rather sensitive to the presence of outliers, and
thus prone to systematics if the sample is not chosen carefully.
If we apply PCA to the whole sample, i.e. with a
single covariance matrix, it will pick a larger signal caused by the
trivial difference 
between galaxies of different types. We would like to focus instead on 
variations within each group, and then among the three
classes. In this way, it is possible to find 
in a robust way systematic differences between the stellar populations
in different groups of galaxies. The BPT classification is taken from the official
galspecExtra SDSS catalog from the MPA-JHU group \citep{Brinchmann:04}.
Regarding quiescent galaxies, we select those with a BPT flag of $-1$, along with 
a threshold on the equivalent width of H$\alpha$ emission, analogously to
\citet{CF:11}.
As explained in section 3, during the calculation of robust covariance
matrices, we remove outliers in each subgroup by culling the samples
at the cost of reducing their sizes. In the 3800-4200\AA\ spectral interval 
the number of galaxies is reduced from 23168 to 17473 (Q); from 10495
to 8025 (SF), and from 3343 to 2620 (A). In the 5000-5400\AA\ spectral interval
the samples are reduced 
from 23168 to 13953 (Q), from 10495 to 6319 (SF), and
from 3343 to 2019 (A). Finally, to ensure that our results
do not depend on the numerical solver, we tested different methods
to diagonalize the covariance matrix. We applied {\sc Python}-based codes
from {\sc Scikit-Learn} ({\tt decomposition.PCA}) and from the {\sc
  LinAlg} package ({\tt eig} and {\tt svd}). All three gave
undistinguishable estimates of the eigenvalues and eigenvectors.

\begin{figure*}
\includegraphics[width=80mm]{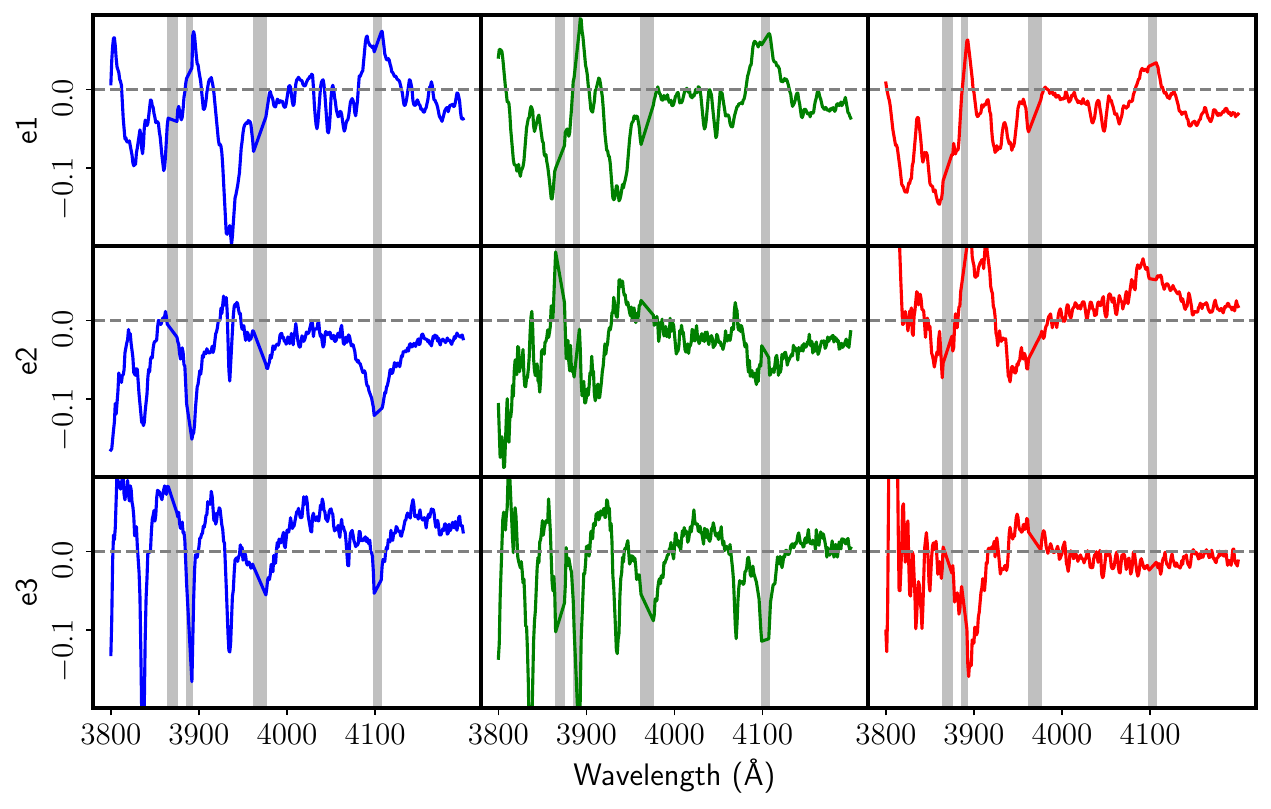}
\includegraphics[width=80mm]{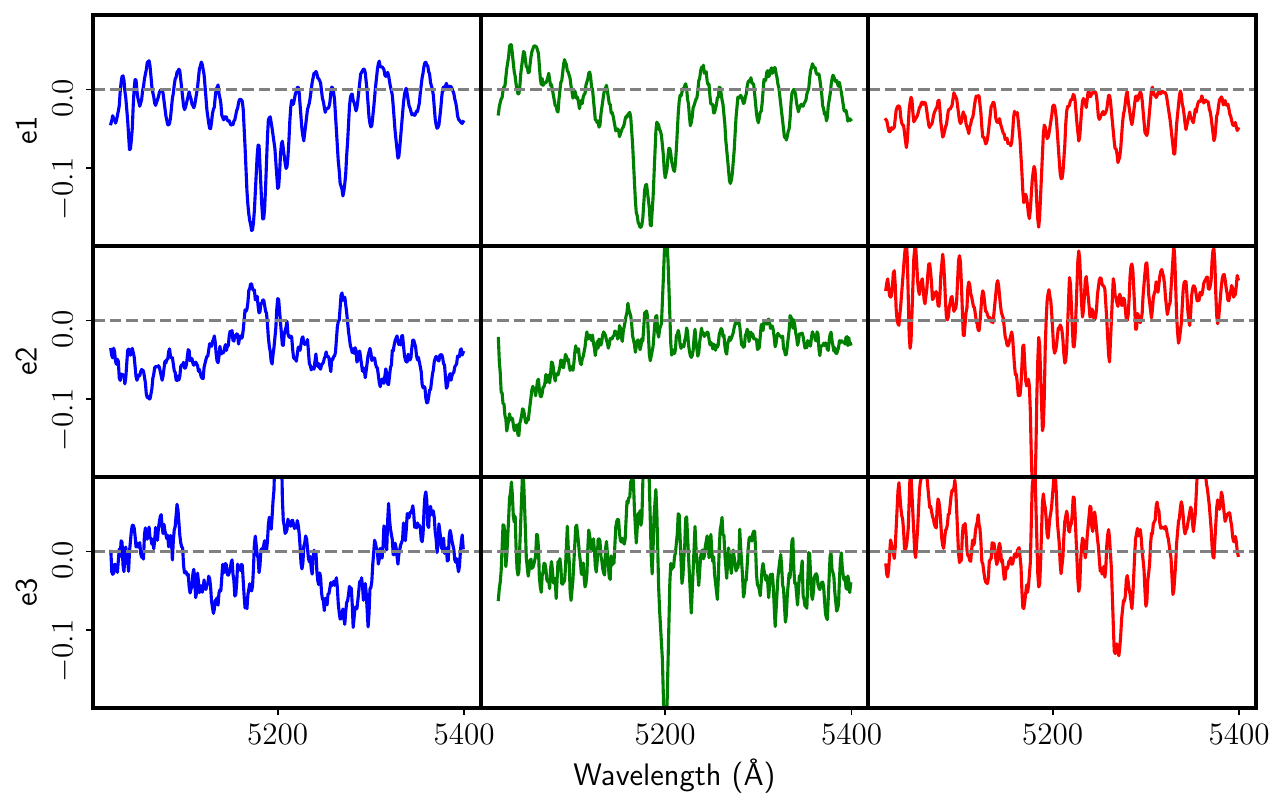}
\caption{Principal components of the three subclasses,
  namely star-forming (blue), AGN (green) and quiescent (red).
  The left (right) panels correspond to the results produced by the
  blue (red) spectral interval. The vertical grey shaded regions
correspond to the intervals removed from the analysis.}
\label{fig:eigenS}
\end{figure*}

Figure \ref{fig:scree} shows the scree plot corresponding to the three
different sets in the blue (left) and red (right) spectral window, colour
coded as labelled. In all cases, the steepness of the scree plots
confirms the dominance of the first few principal components. It is
worth noting that in both choices of wavelength window, AGN spectra
have a higher fraction of total variance in the first component with
respect to SF or Q.  We also note
the contrast in these scree plots with analyses based on full spectra
(i.e. including the continuum), where, for instance, early-type
galaxies include over 90\% of total variance in the first few 
components \citep[see, e.g.][]{Rogers:07}. Removing the
continuum makes the data less compressible. In our sample, performing
the same analysis in the blue interval on data without continuum
subtraction results in a fractional variance of the first principal
component of 73\% (AGN); 84\% (SF) and 36\% (Q),
in marked constrast to our adopted, continuum-substracted, spectra,
for which the contribution of the first component is 
12\% (AGN); 8\% (SF) and 3\% (Q) of total variance. 
This confirms the continuum encodes a large amount of variance, or information.
However, our approach is aimed at minimising all potential
systematics, and dust can produce substantial entanglement with
respect to variations in the underlying stellar populations.  In
principle, the higher order components will be more severely affected by noise,
but this interpretation is not straightforward, and truncating the PCA
decomposition to denoise spectra is a far from trivial task.  The blue
interval keeps a higher ratio of the variance in the first component
($\sim$10\%) with respect to the red interval ($\sim$5\%), which
is perhaps indicative of a higher level of correlatedness at bluer
wavelengths (i.e. higher compressibility). In absolute terms, the blue
interval has substantially higher variance: for instance AGN spectra
have absolute variance of 0.357 in the first component for the blue interval,
and 0.036 in the red interval. This paper focuses on the
lowest order components of variance, so we will restrict the analysis
to the projections of the data on to the first three principal components.

The first three eigenspectra of the SF (blue), AGN (green), and Q
(red) galaxies are shown in Fig.~\ref{fig:eigenS} in both spectral
windows (left/right panels).  The vertical grey shaded regions
correspond to the intervals removed from the analysis. We note that the covariance matrix
is sign invariant, i.e.  a change in the sign of a principal component
(and thus its projections) does not affect the matrix. Therefore, the
comparisons between the three covariance matrices can be affected by a
sign change. In order to mitigate this issue, we enforce a positive sign for
the median of the projections of all galaxies of the corresponding type.
When restricting to the first three principal components,
this means the data are statistically constrained to the first octant of the latent
space spanned by \{PC1, PC2, PC3\}.

\section{Projecting the data on to the principal components}
\label{Sec:Proj}

Once we have set the three eigenspectra with the highest variance as those
that encapsulate most of the information regarding absorption lines, we
project the SDSS spectra on to those, separated with respect to their
classification as SF, AGN or Q.
For instance, principal component number one for the $j$th galaxy is defined as:
\begin{equation}
{\rm PC1}_{j} = \Phi_{j}\cdot\hat{\bm{e}}_1 = \sum\limits_{i=1}^N\Phi_j(\lambda_{i})\hat{\bm{e}}_1(\lambda_{i}),
\label{eq:PC1}
\end{equation}
where N is the total number of wavelength bins in the spectrum.
We note that we only present the
projected sample that have not been  rejected by the condition
described in eq.~\ref{eq:4sigma},
although the results do not vary significantly if the full sample
is considered instead.
Each spectrum is thus simplified 
by three ``coordinates'', namely PC1, PC2 and PC3, and can be visualized in a three
dimensional latent space. We emphasize that the results are obtained from 
continuum subtracted spectra, rejecting regions dominated by nebular emission
lines, so that the analysis does not depend on the dust or gas components.

\begin{figure*}
\includegraphics[width=80mm]{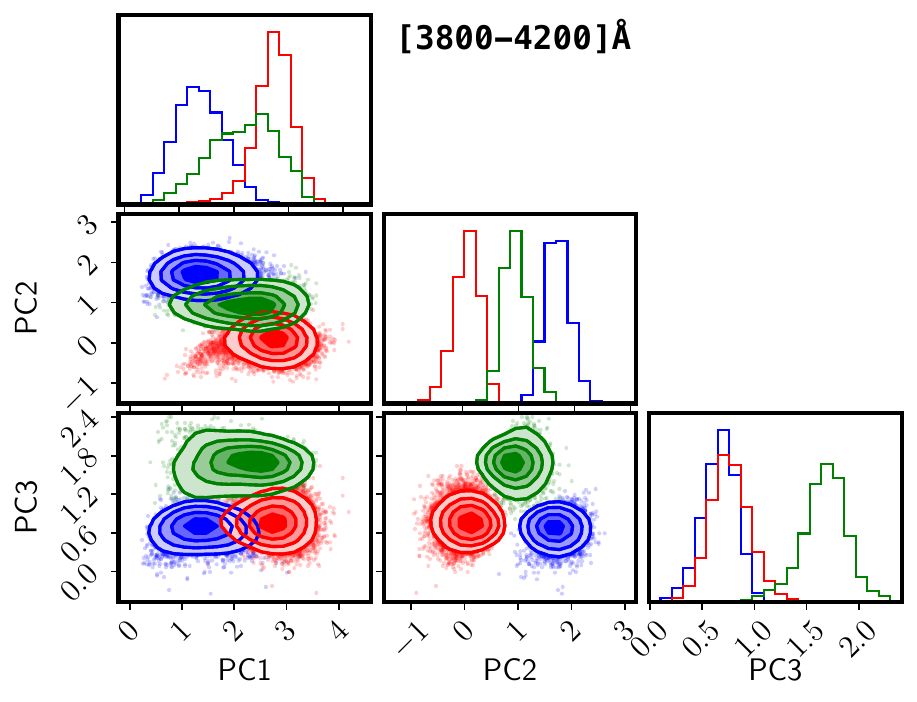}
\includegraphics[width=80mm]{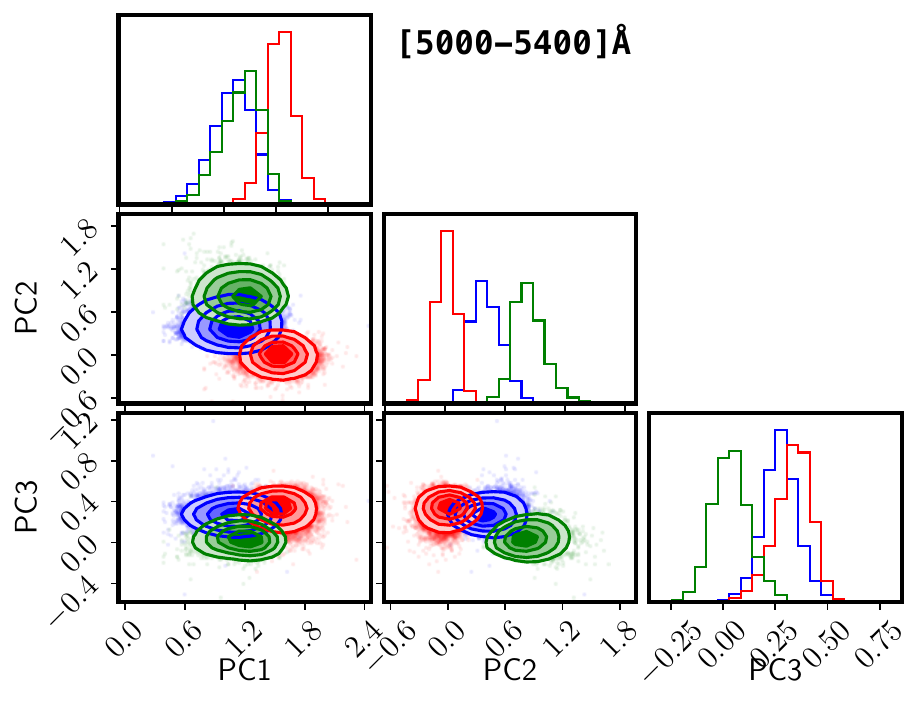}
\caption{Distribution of the projections of the galaxy spectra on to the first three principal
  components, colour coded as star-forming (blue), AGN (green) and quiescent (red).The left (right) panels correspond to the results produced by the blue (red) spectral interval. The contours engulf 25, 50, 75 and 90\% of each subsample.
  See Fig.~\ref{fig:animation1} as complement to this figure.}
\label{fig:corner}
\end{figure*}

\begin{figure}
  \centering
  \includegraphics[width=80mm]{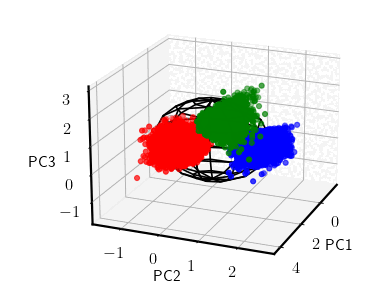}
  \includegraphics[width=80mm]{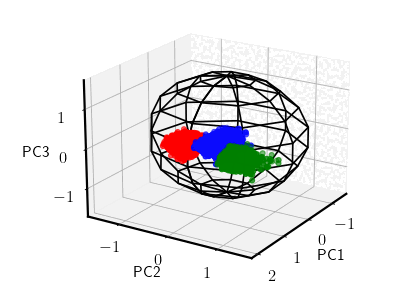}
  \caption{Distribution of the projections of
    a representative sample of galaxy spectra on to the first three principal
    components. The data are colour coded as star-forming (blue), AGN (green) and quiescent (red).
    The top (bottom) panels correspond to the results produced by the blue (red) spectral interval.
    The framed structure represents a sphere with radius 1, and is only shown for reference.
    An animated version of the figures can be found at
    \href{https://cloud.iac.es/index.php/s/firKQ2BFkC7xzSf}{this link} (for the blue interval) and
    \href{https://cloud.iac.es/index.php/s/deniTo6SgpiBsR4}{this link} (for the red interval).}
  \label{fig:animation1}
\end{figure}

\begin{figure*}
\includegraphics[width=85mm]{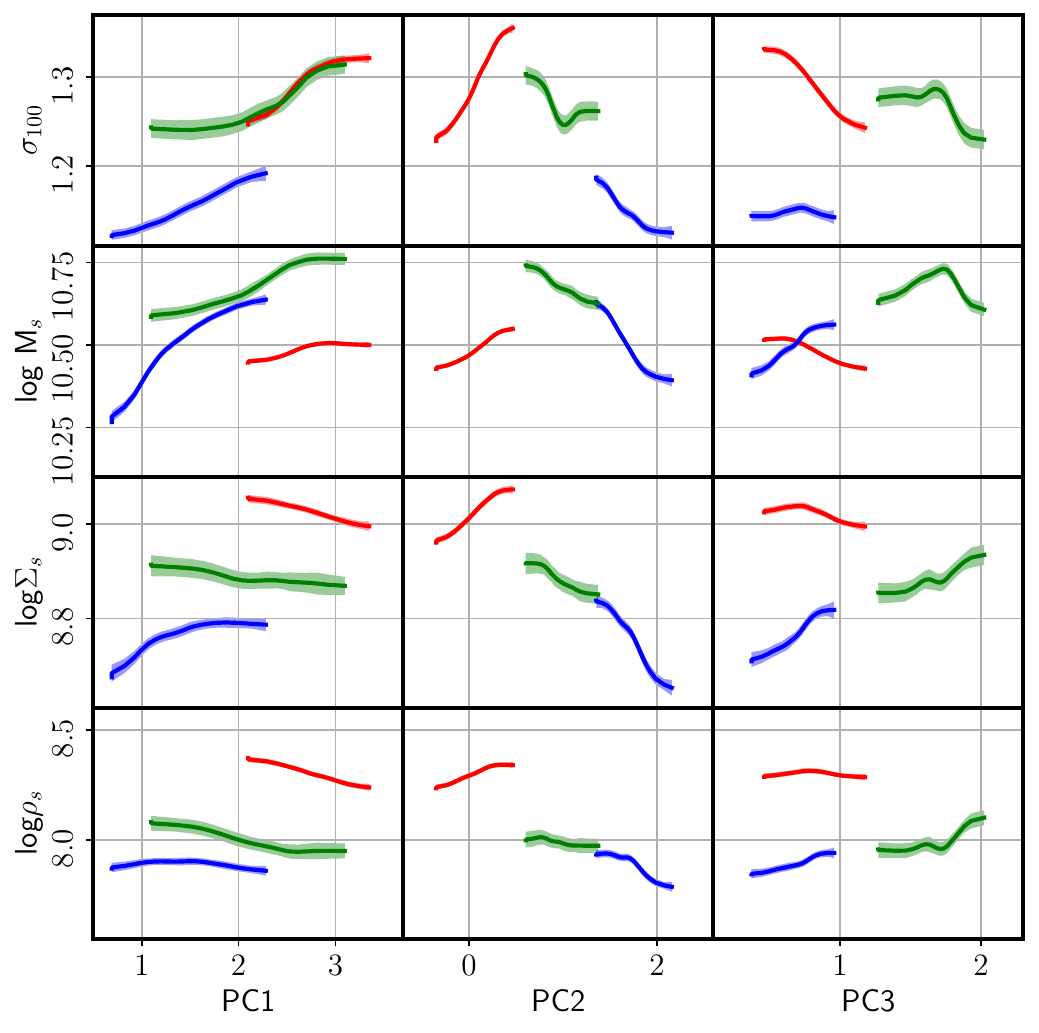}
\includegraphics[width=85mm]{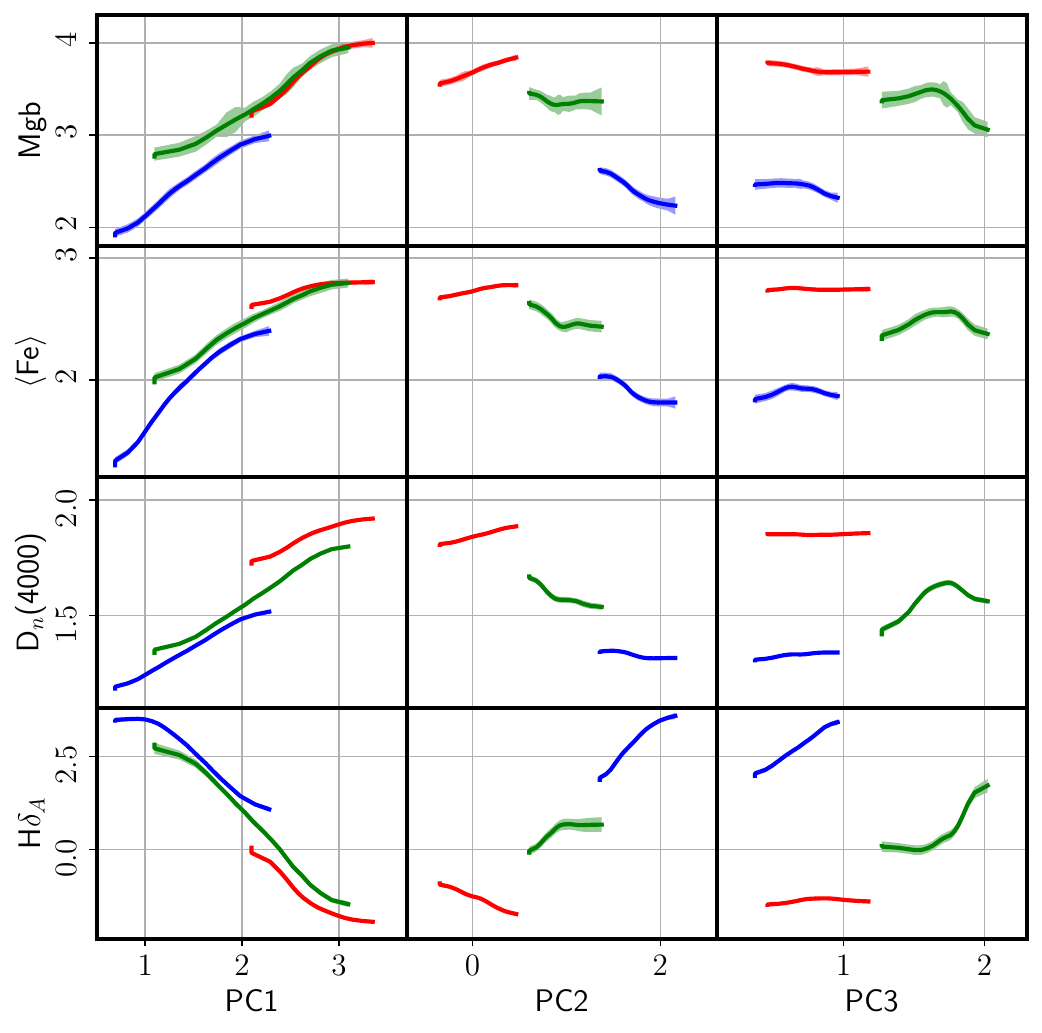}
\caption{Correlation of the projections on to the first three
  principal components with observable features of galaxies in the
  blue spectral range (3800-4200 {\AA}). In each panel, the data are shown separately
  for quiescent (red), star-forming (blue) and AGN (green), as a
  running median towards increasing values of the PC1, PC2, PC3
  projections. The shaded regions mark the 1$\sigma$ uncertainty of the median.
  See text for details.}
\label{fig:Corr1}
\end{figure*}

Fig.~\ref{fig:corner} shows the latent space of the SDSS galaxy
spectra, in blue (SF), green (AGN) and red (Q), in both spectral
windows: blue (left) and red (right). We use the publicly available
{\sc Python} code of \citet{corner} to produce a visual representation
of the distributions in the three components with the so-called
``corner'' plots. The 2D projections are shown as a
colour scale mapping the number density of points at a given position.
The most salient feature is that the spectra are projected in
separate regions of their respective latent spaces, regarding their nebular
emission properties.
We emphasize that although the figure shows projections onto different
eigenvectors, depending on their nebular activity classification, the underlying
data concern one and the same component, namely the stellar populations of
the galaxies (after careful removal of features affected by dust or
ionised gas). Therefore Fig.~\ref{fig:corner} is not a disjoint
comparison of PCA projections, and the eigenvectors of the three groups
are not completely independent. The three sets of eigenvectors reflect, instead,
the typical stellar populations found in each group. The separation between the three
subgroups found in the figure is reminiscent of the bimodal
distribution into blue cloud (i.e. star forming) and red sequence
(i.e. quiescent) galaxies \citep[e.g.][]{Strateva:01,Baldry:04}, with
AGN preferentially populating the green valley \citep[see,
  e.g.][]{Schawinski:12,Salim:14,Angthopo:19}.  The separation is more
pronounced in the blue interval, where both PC1 and PC2 show clear
separation between the three groups, with AGN located in between the
other two. PC3 puts SF and Q galaxies together, and AGN standing out
with higher values. In the red interval, the distributions also appear
separately when projected on to PC2; whereas PC1 appears to
``isolate'' Q galaxies and PC3 sets AGN aside.  To help visualize the
distribution of principal component projections, we show in
Fig.~\ref{fig:animation1} a rotating animation of the 3-dimensional
latent space (with a link provided in the caption).
Note that while PCA is independently applied to three
different covariance matrices, which results in different
eigenvectors, their clustering is significant even considering the
different orientations. Given the input data lack, by construction,
emission lines of even continuum, the clustering is representative of
the variations in the stellar population content.  These figures also 
illustrate one of the biggest problems encountered in a data-driven,
multivariate approach: it is very difficult to ascribe such
differences to specific physical/observational factors. The
eigenvectors in Fig.~\ref{fig:eigenS} have some physical features but
they cannot be associated, say, to specific populations, or phases of
evolution. Therefore, in order to gain more insight of the meaning of
these projections, we show in Figs.~\ref{fig:Corr1} and
\ref{fig:Corr2} the way these projections depend on a number of
observable properties.  In each panel, we separately sort each
subsample (Q, SF and AGN) in increasing order of the PC1, PC2 or PC3
projection, and show the running median -- choosing bins that
encompass 1/20th of each subset, therefore comprising from 100 to over
800 galaxies per bin -- along with the 1$\sigma$ uncertainty of the
median, shown as a shaded region.  We follow the same colour coding as
above. The panels on the left trace more general observables, from top
to bottom: stellar velocity dispersion in units of 100\,km/s; stellar
mass (measured as $\log\,$M/M$_\odot$); surface stellar mass
($\log\Sigma_s$); and volume stellar mass ($\log\rho_s$). The last two
are defined as $\Sigma_s\equiv$M$_s$/R$_P^2$, and
$\rho_s\equiv$M$_s$/R$_P^3$, respectively, where M$_s$ is the total
stellar mass in solar units and R$_P$ is the Petrosian radius in the
SDSS-$r$ band, measured as a projected physical distance at the
redshift of the galaxy, in kpc.  The panels on the right show
spectroscopic measurements, from top to bottom: Mgb; average
Fe=0.5(Fe5270+Fe5335) \citep{Trager:98}; D$_n$(4000) \citep[as defined
  in][]{Balogh:99}, and H$\delta_A$ \citep{WO:97}.  These estimates
are all derived from the official SDSS catalogues.  The results are
presented for the blue interval (Fig.~\ref{fig:Corr1}) and the red
interval (Fig.~\ref{fig:Corr2}).  We emphasize that these projections
concern continuum-subtracted data without any contribution from
emission lines. Therefore, the associated variance is independent of
the dust and gas components.

The figure shows a strong correlation of PC1 with all spectral
observables, unsurprisingly, given that PCA produces the components
based on the variance of the data, which reduces to the absorption
lines. However, it is rather unexpected to find the much weaker
dependence of the higher order projections PC2 and PC3. The scree
plot indicates a substantial fraction of total variance in these
components, nevertheless, they appear not to correlate so strongly
with the spectral data. The red interval produces similar results,
but Mgb and $\langle$Fe$\rangle$ feature a sizeable correlation
with PC2 and PC3, with respect to the blue interval. It is a well-known
fact that all spectral indicators are deeply entangled, i.e. the
so-called age-metallicity degeneracy \citep{W:94, FCS:99} has a
deeper version when considering the covariance among different
spectral indicators \citep[see figs.~10 and 11 of][]{InfoPop}.
Regarding the general observables (left panels), both blue
and red intervals display a strong correlation with velocity
dispersion. The relation between $\sigma$ and PC2 is stronger in quiescent
galaxies, and especially in the red interval. These galaxies tend to
have stronger absorption features (see Fig.~\ref{fig:SigPlot}), and the Mgb--Fe region is 
optimal to measure velocity dispersion. However, it is noteworthy that
this signal (roughly the width of the absorption lines) is
listed by PCA as the second largest variance, with the depth of these
features ranking higher -- i.e. the trend of PC1 with the line strengths.
Stellar mass and surface/volume densities also
correlate with the projections, but we conclude that this trend
is inherited by the relation with $\sigma$, as this parameter
strongly correlates with both.
We should emphasize here that this sample extends over a relatively
small range of velocity dispersion -- on purpose, to avoid
``distracting'' the variance with the well-known age-mass and
metallicity-mass correlations \cite[e.g.][]{Gallazzi:14}.
Nevertheless, the data clearly display a version of these scaling laws
over the narrower range of mass and velocity dispersion.

It is also worth mentioning that these trends are followed in
different ways in Q, SF and AGN spectra, especially in the higher
order (i.e. lower variance) projections. The most evident discrepancy
between these subclasses is average age, and perhaps chemical
composition, but more subtle differences might be due to details of
the underlying stellar populations.

\subsection{Projection of Simple Stellar Populations}

\begin{figure*}
\includegraphics[width=85mm]{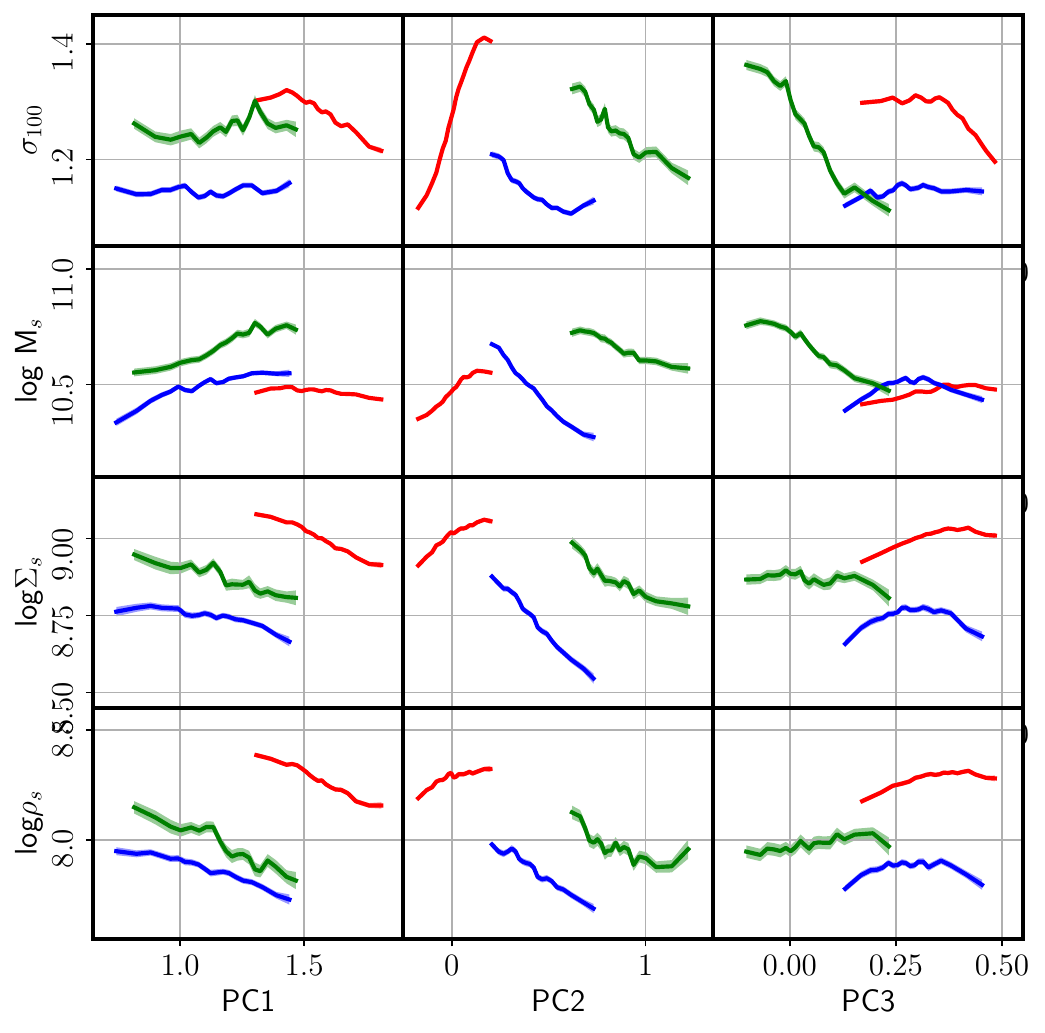}
\includegraphics[width=85mm]{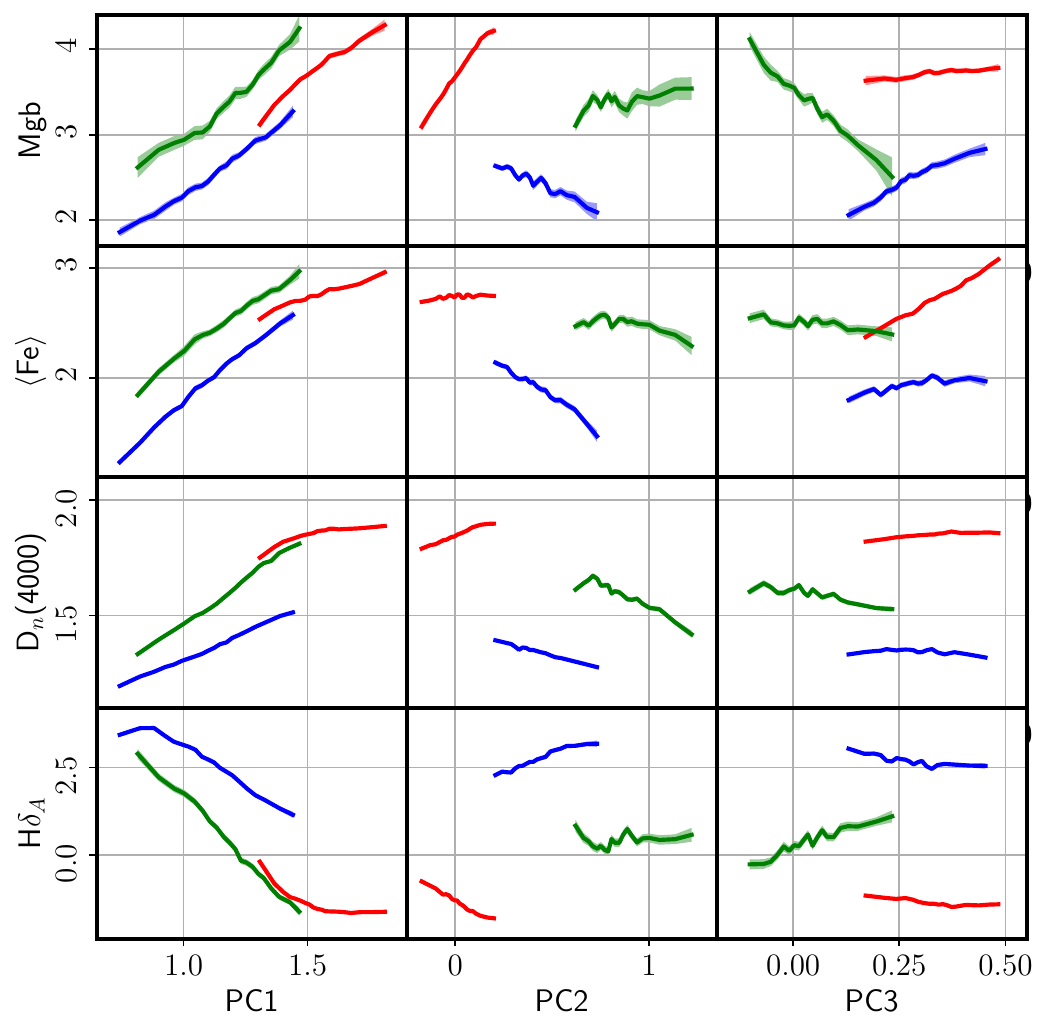}
\caption{Equivalent of Fig.~\ref{fig:Corr1} for the analysis of
  the red spectral interval (5000-5400\AA).}
\label{fig:Corr2}
\end{figure*}

\begin{figure*}
\includegraphics[width=85mm]{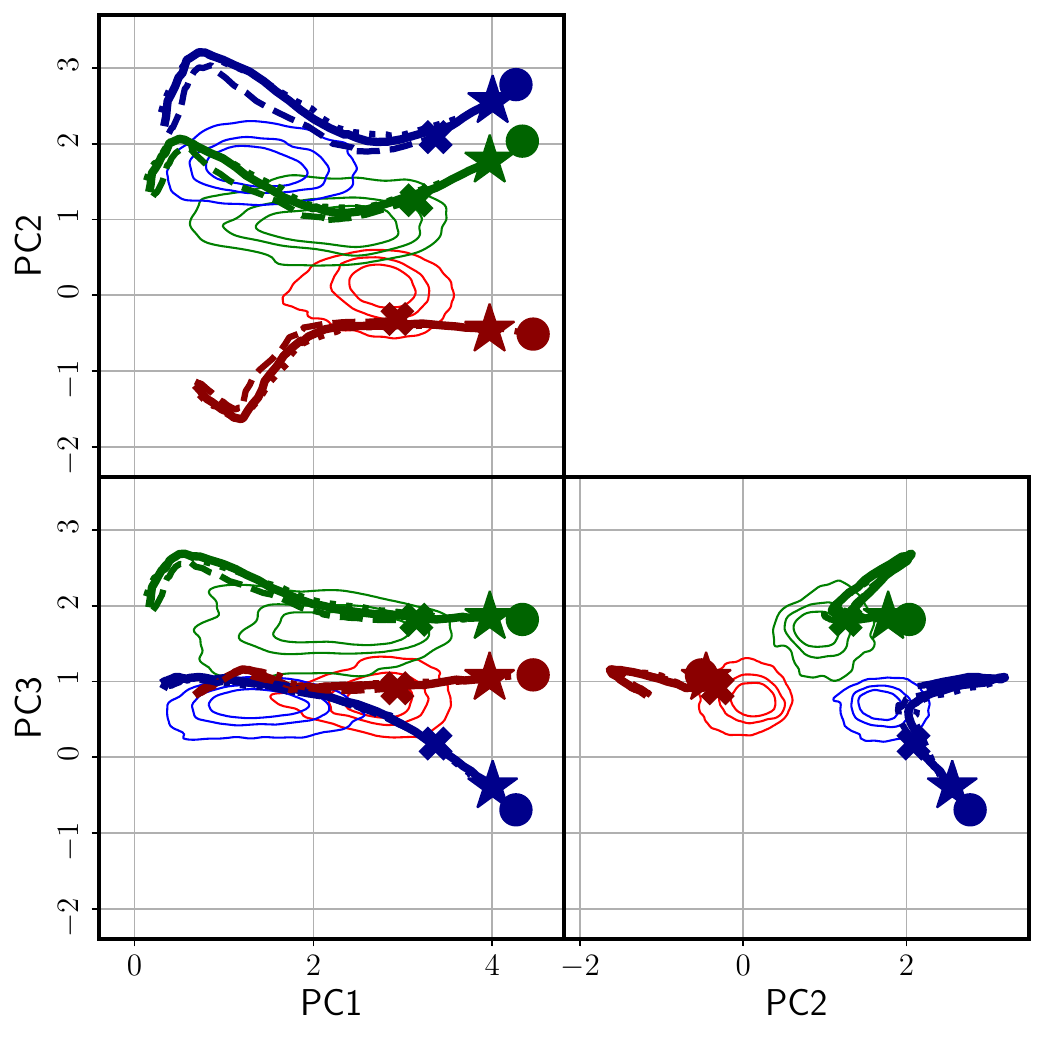}
\includegraphics[width=85mm]{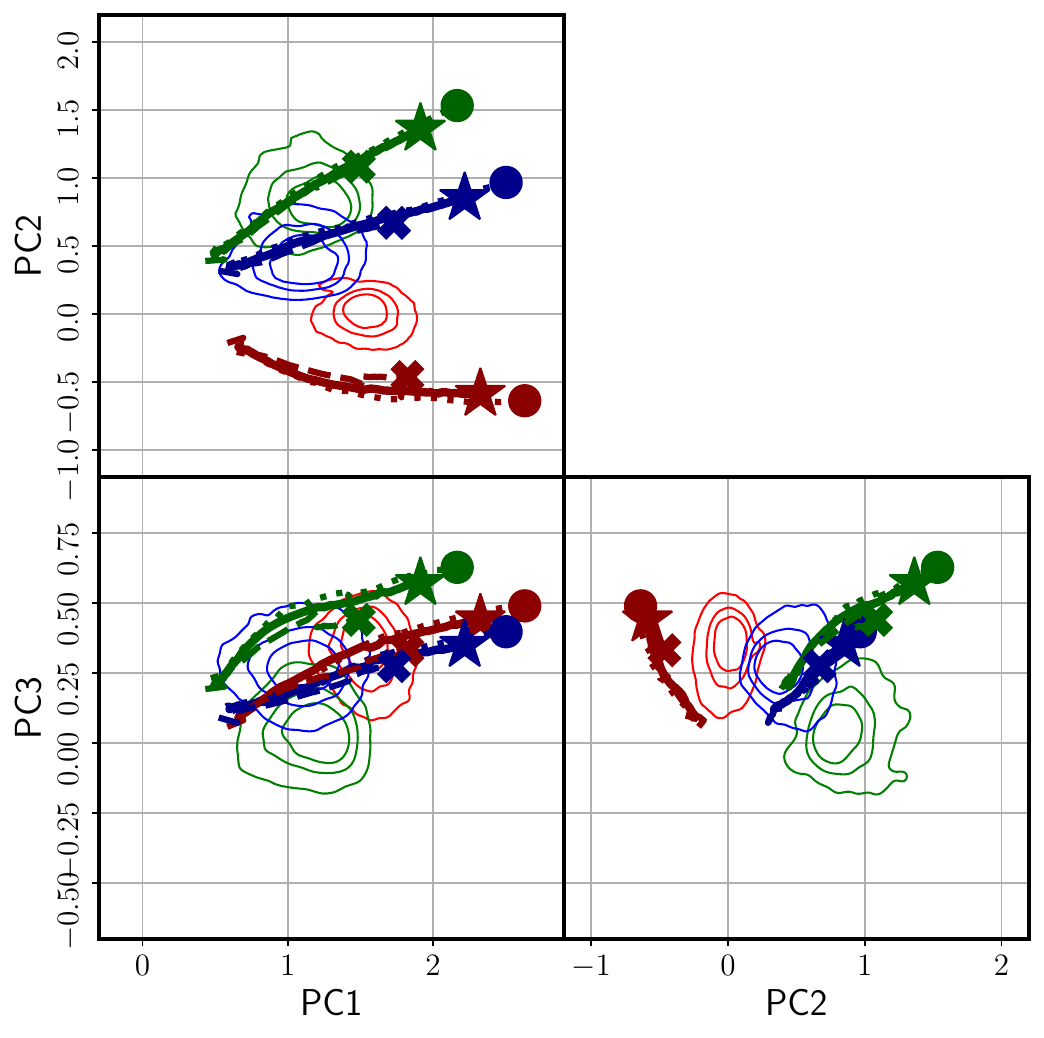}
\caption{Projection of simple stellar populations from the MIUSCAT
  models \citep{MIUSCAT} on to the first three principal components.
  The age ranges from 0.1 to 14\,Gyr, the oldest one represented by
  a symbol, with metallicity [Z/H]=$-0.4$ (dashed), 0.0 (solid) and
  $+0.22$ (dotted). For reference, the number density of the
  SDSS sample is shown as contours engulfing 50, 75 and 95\% of each subset.
  The colour coding is the same as in Fig.~\ref{fig:corner}. The left (right) panels correspond to the results produced by the blue (red) spectral interval.}
\label{fig:SSPtracks}
\end{figure*}

\begin{figure}
  \centering
  \includegraphics[width=80mm]{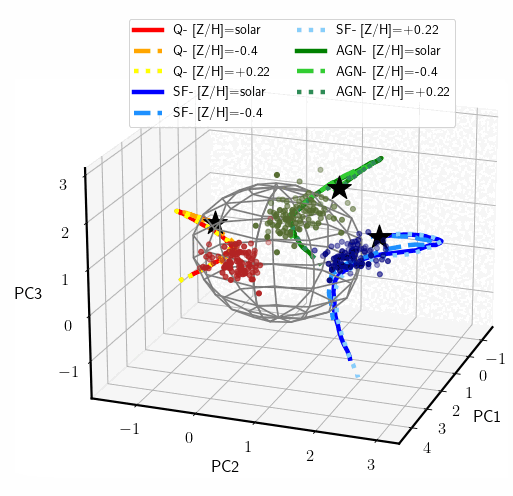}
  \includegraphics[width=80mm]{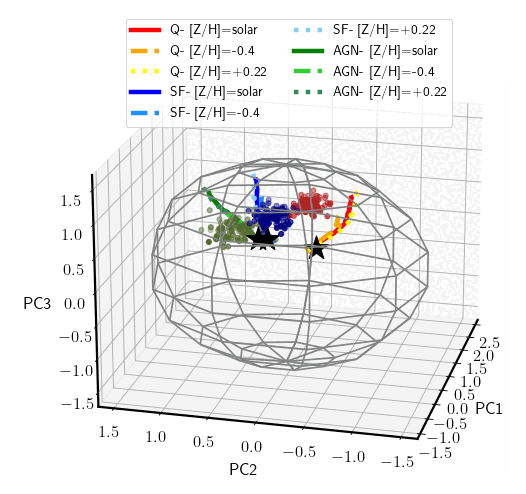}
  \caption{Distribution of the projections of a representative
    sample from each of the three subsets onto the first three
    principal components, along with the corresponding projections
    of SSPs from population synthesis models. The data are colour
    coded as star-forming (blue), AGN (green) and quiescent (red).
    The results for the blue (red) spectral interval is shown in
    the top (bottom) panels.  The framed structure represents a
    sphere with radius 1, and is only shown for reference.
    An animated version of the figures can be found at 
    \href{https://cloud.iac.es/index.php/s/HDcQftPJsYFEoxG}{this link} (for the blue interval) and
    \href{https://cloud.iac.es/index.php/s/7Q2NEBnEQ3PWAYM}{this link} (for the red interval).}
  \label{fig:animation2}
\end{figure}

The trends of the principal component projections shown in
Figs.~\ref{fig:Corr1} and \ref{fig:Corr2} have been obtained in a model-independent
way. Therefore, they are free of the inherent systematics caused by
comparisons with models that suffer from degeneracies. However, we
cannot give physical meaning to those. We now turn to stellar
population synthesis, with the aim of translating those trends into
parameters that describe, to the lowest order, the properties
of the stellar component. We firstly produce simple stellar populations
from the MIUSCAT models \citep{MIUSCAT} for a range of
ages (0.1$<$t$_{\rm SSP}<$14\,Gyr), at three
different metallicities: [Z/H]=$\{-0.4,0,+0.22\}$,
with a fixed stellar initial mass function \citep{KroupaU:01}. The SSP
spectra are convolved to the mean velocity dispersion of the
observed galaxies, and resampled to the same wavelength binning.
Moreover, following the same methodology as with the SDSS data,
we remove the pseudo-continuum. While these population synthesis
models do not include nebular emission from diffuse gas, we
also remove the regions contaminated
by emission, for a consistent analysis.
Finally, we project the processed spectra onto the
principal components obtained from the analysis, showing the
synthetic data as three separate age tracks in Fig.~\ref{fig:SSPtracks},
one for each choice of metallicity (solid lines for solar metallicity,
dotted for super-solar, and dashed for subsolar metallicity).
For reference, the oldest population is shown as a symbol
(star, circle and cross for solar, super-solar and subsolar metallicity,
respectively). Note that there are three separate sets of tracks
as we project the SSP spectra on to the principal components
from the Q (red), SF (blue) and AGN (green) covariance matrices.
For reference,
we also show as a contour plot, the number density of observed
spectra (each contour encompases, from the inside-out, 50, 75
and 95\% of the data). The left (right) panel corresponds to the
analysis in the blue (red) interval. Fig.~\ref{fig:animation2} is an
animated 3D version of the projections shown in Fig .~\ref{fig:SSPtracks},
using a representative sample of 100 data points, selected randomly
from each subset (a link to the animated figures is provided in the caption).

The most noticeable result of this comparison is the strong age
dependence of PC1, in both spectral intervals.  In the red interval
PC2 and PC3 also have a monotonic trend with age, whereas in the blue
interval the higher order components have more subtle variations, with
PC2 changing only for younger ages (similar to, e.g. Balmer
absorption), and PC3 has a substantial correlation with age only
for the SF subsample.

Regarding metallicity, these projections illustrate in a remarkable
way the fundamental nature of the age-metallicity degeneracy. In this
3D latent space, the effect of changing metallicity is equivalent to a
change in age, as the tracks at different metallicity simply ``slide''
with respect to each other, in the expected direction that increasing
metallicity produces an equivalent effect as an increase of age. We
emphasize the importance of this result, as the traditional case is
based on photometric and spectroscopic observables \citep[e.g.,][]{FCS:99}, whereas this
result depends at the most fundamental level on the intrinsic variance
of stellar population spectra. Finally, we note the acceptable
agreement between synthetic spectra and real galaxies, except for the
case of quiescent galaxies in the red interval, where the synthetic
tracks do not overlap the bulk of the observations.  As this region is
dominated by the Mgb--Fe complex, we suggest this discrepancy may be
due to non-solar abundance ratios, more prevalent in quiescent
galaxies \citep[e.g.][]{Trager:00,Kuntschner:10}.

\subsection{Spectral fitting of projected data}

\begin{figure*}
\includegraphics[width=.05\linewidth]{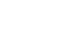}
\includegraphics[width=.3\linewidth]{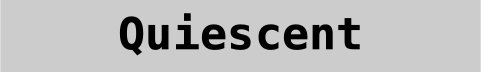}
\includegraphics[width=.3\linewidth]{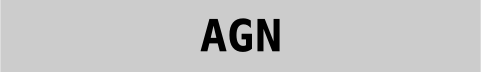}
\includegraphics[width=.3\linewidth]{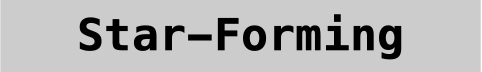}
\includegraphics[width=.05\linewidth]{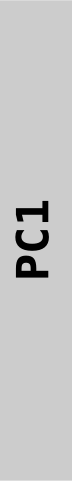}
\includegraphics[width=.3\linewidth]{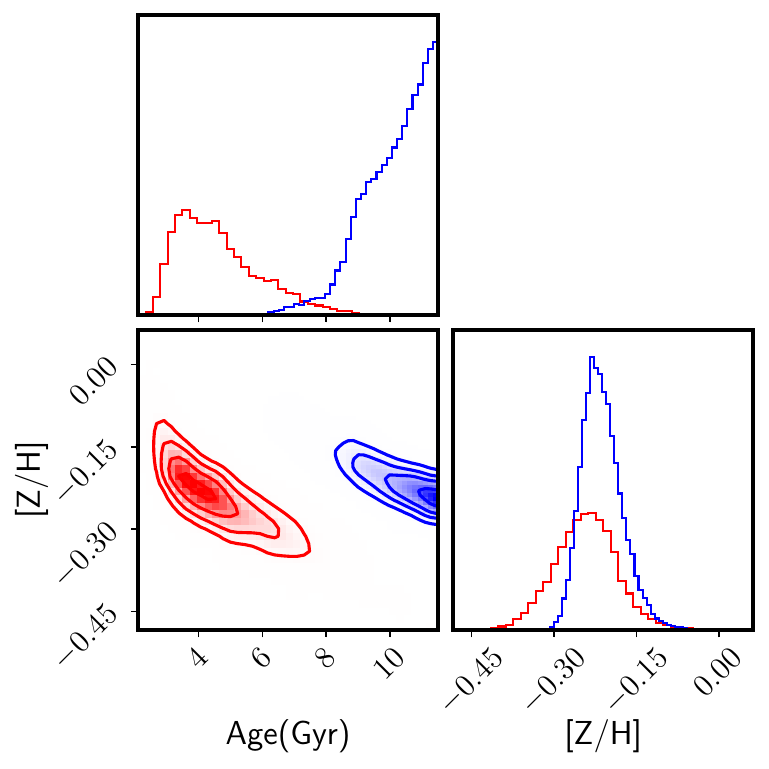}
\includegraphics[width=.3\linewidth]{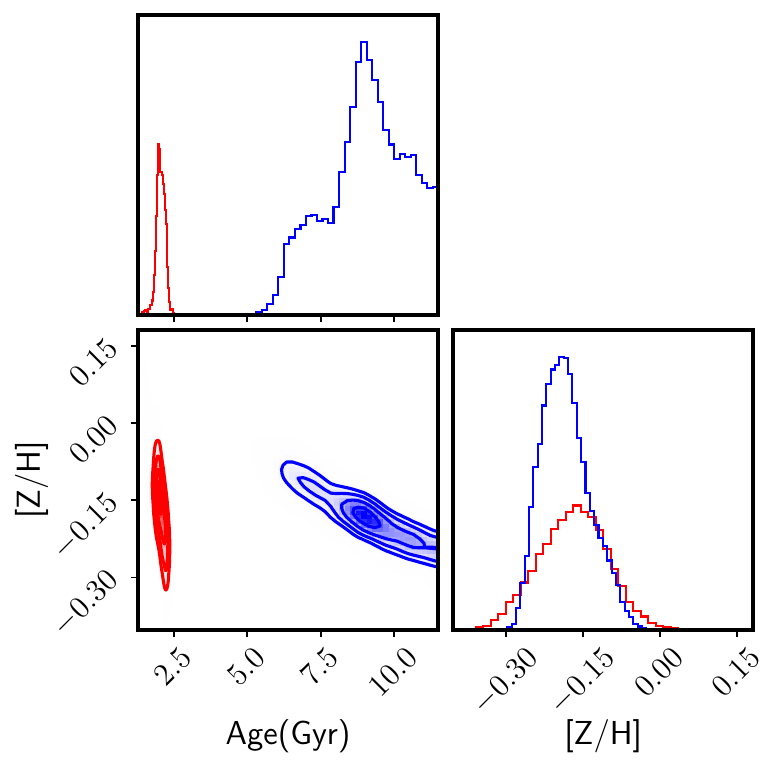}
\includegraphics[width=.3\linewidth]{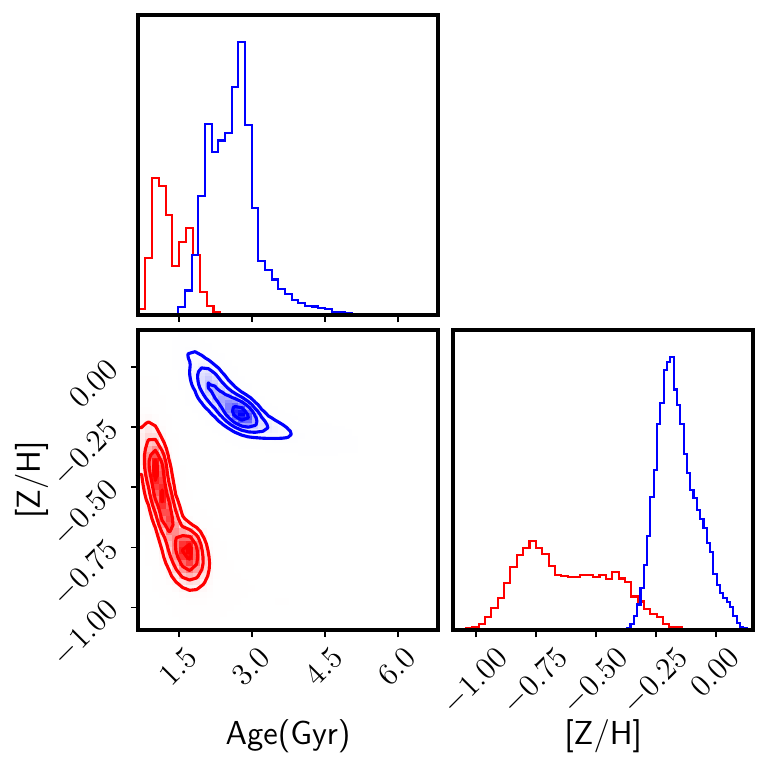}
\includegraphics[width=.05\linewidth]{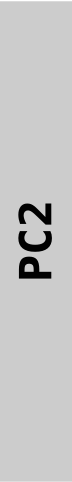}
\includegraphics[width=.3\linewidth]{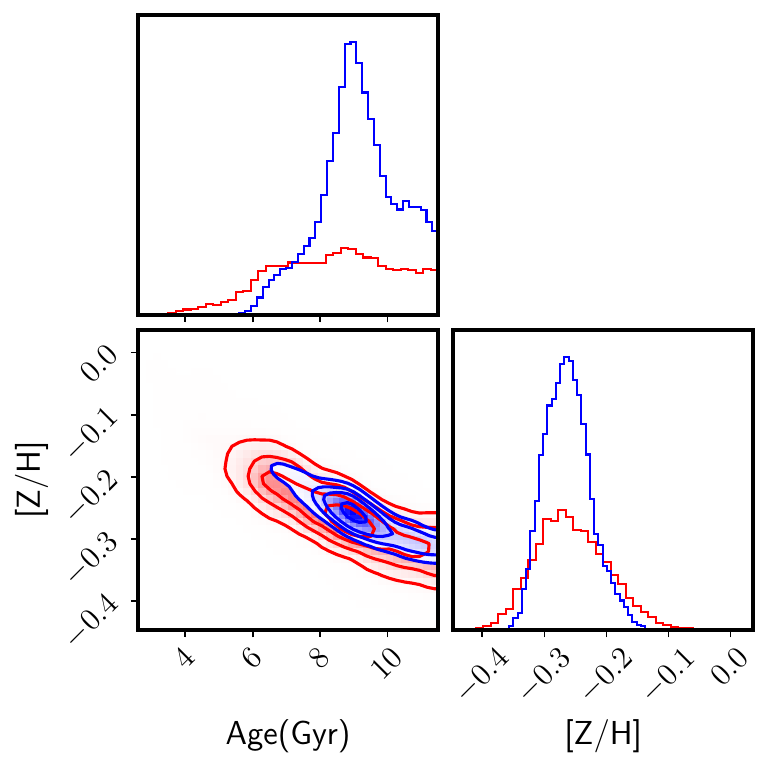}
\includegraphics[width=.3\linewidth]{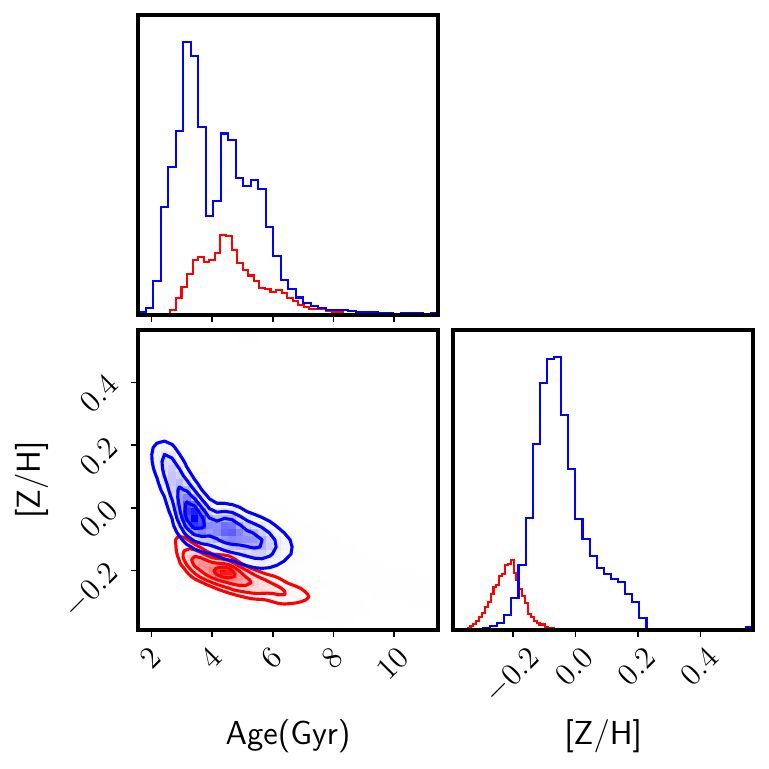}
\includegraphics[width=.3\linewidth]{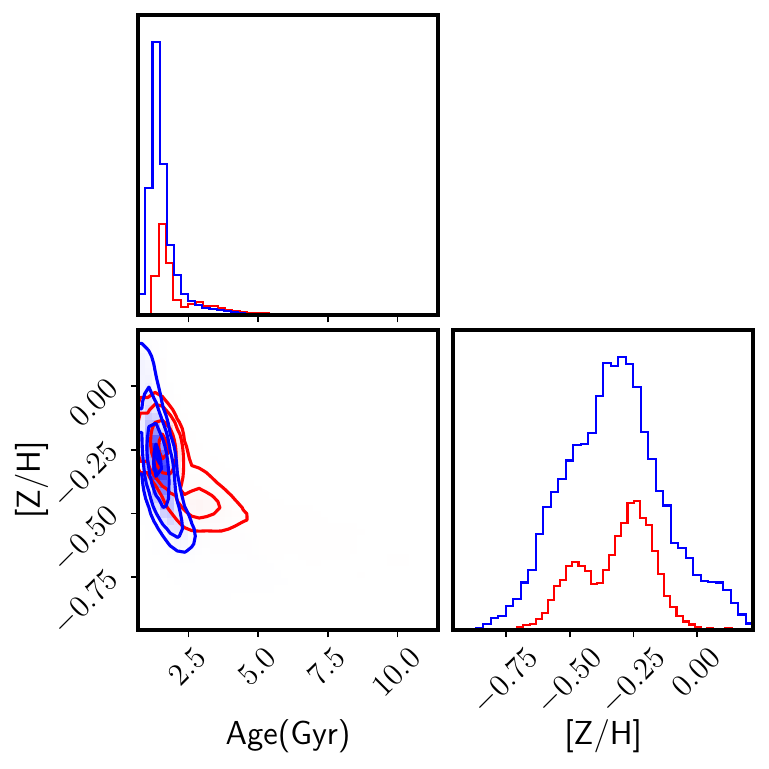}
\includegraphics[width=.05\linewidth]{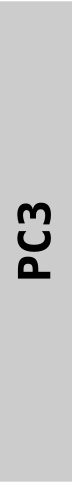}
\includegraphics[width=.3\linewidth]{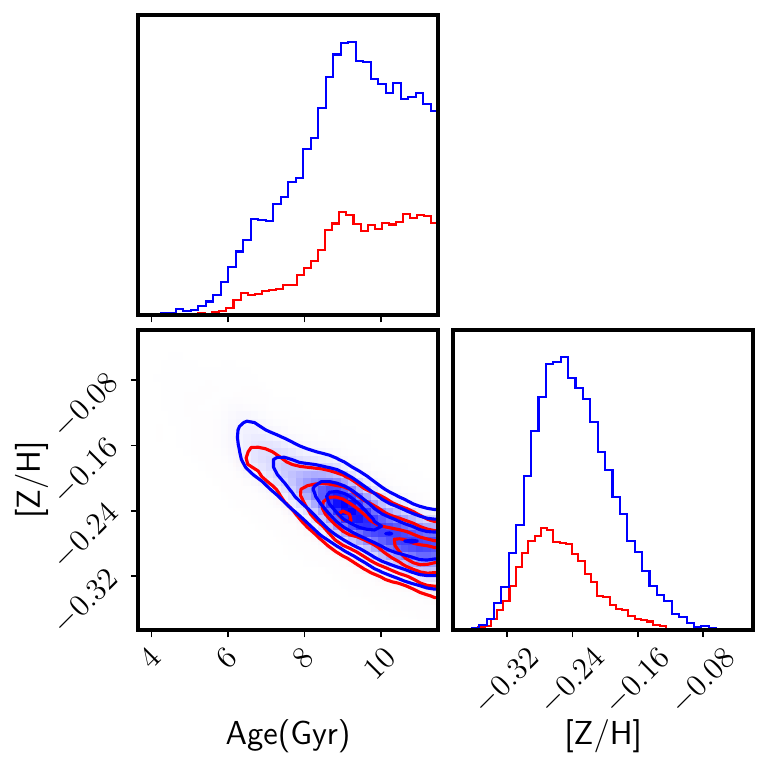}
\includegraphics[width=.3\linewidth]{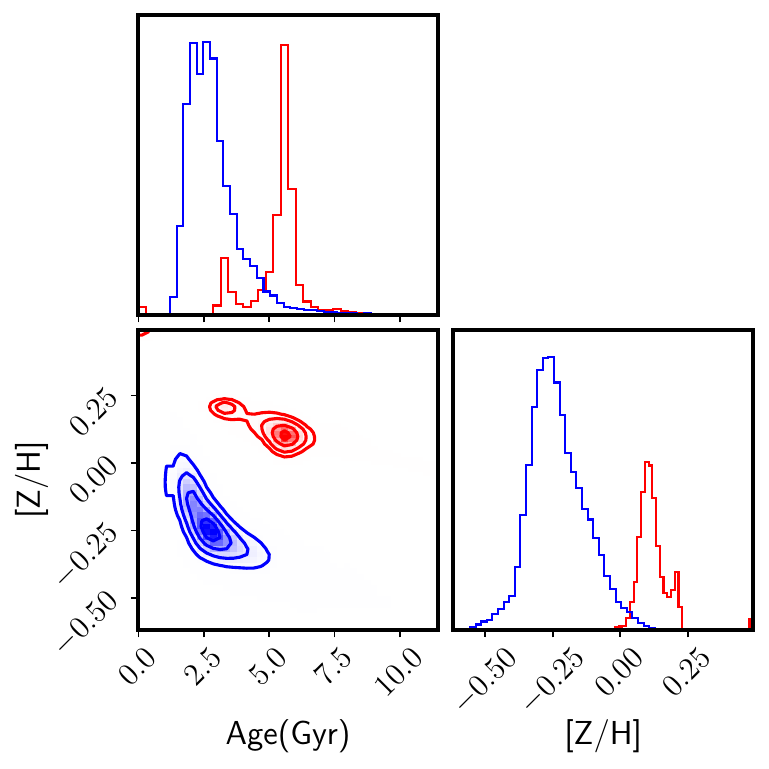}
\includegraphics[width=.3\linewidth]{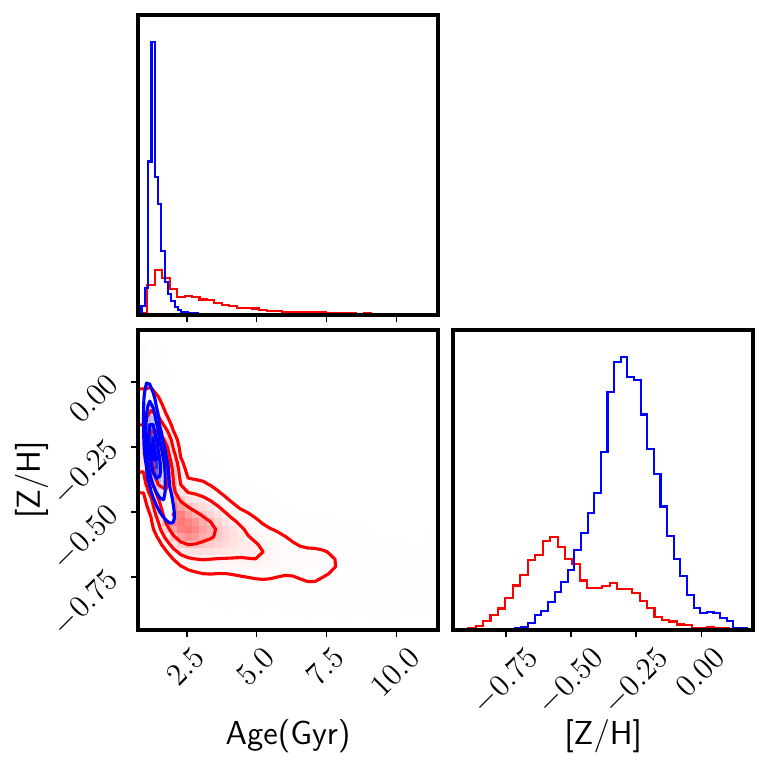}
\caption{Confidence levels of the SSP-equivalent age (in Gyr) and metallicity ([Z/H]) 
  obtained from fitting the spectra (continuum subtracted) with
  the MIUSCAT population models \citep{MIUSCAT}. We fit 
  the stacked spectra with the lowest (10 percentile, red)
  and highest (90 percentile, blue) value of PC1, PC2 and PC3 (from top to bottom),
  for Q (left column), AGN (middle column) and SF galaxies (right column).
  The principal components obtained correspond to the covariance
  defined in the blue interval.
}
\label{fig:MCMC}
\end{figure*}

Building upon the previous comparison of synthetic data projected
on the latent space generated by the real spectra, we now make a
more standard analysis by stacking spectra within a given
(SF/AGN/Q) subclass regarding their projections. For each choice of the spectral interval and
nebular emission type, we produce two stacked spectra, combining the data from galaxies in
the 10th and 90th percentile of the distribution of each principal component.
The resulting stacks are
then compared with SSPs using a standard method based on the
$\chi^2$ statistic, fitting the continuum-subtracted data -- for consistency
with our analysis, the
continuum does not play any role in the fitting -- and
adopting the standard likelihood
${\cal L}(t,[Z/H])\propto\exp(-\chi^2(t,[Z/H])/2)$,
where the only two parameters considered are the age (exploring the 0.01-11.5\,Gyr
range) and total
metallicity (between [Z/H]=-2.3 and +0.2) of the
SSP \citep[keeping the stellar initial mass function of][]{KroupaU:01}. 
We adopt the {\sc Python}-based MCMC solver {\tt emcee} \citep{emcee} to
produce the confidence levels of the fits, shown in
Fig.~\ref{fig:MCMC}. The contours are shown at the 1, 2, 3, and 4\,$\sigma$ 
levels, with the lowest PC projections in red (10th percentile) and the
highest in blue (90th percentile).  The columns, from left to right,
represent Q, AGN and SF spectra, and the rows, from top to bottom,
correspond to the distribution of PC1, PC2 and PC3 projections. We
use the same population synthesis models as in
Fig.~\ref{fig:SSPtracks}.

Consistently with the projected tracks, we find that the
highest/lowest values of PC1 correspond to the oldest/youngest
galaxies, respectively, in all three subgroups, with the average
populations being, unsurprisingly,  younger in the sequence  
Q$\rightarrow$AGN$\rightarrow$SF. No significant difference is found
with respect to metallicity in Q and AGN, whereas SF stacks also show
different metallicities in an opposite direction to the
age-metallicity degeneracy (the latter tracing the elongated confidence levels,
from top left to bottom right
in these plots). This may be caused by the fact that the degeneracy
is usually more entangled in older populations.
PC2 produces mixed results: Q spectra do not show any appreciable
difference, AGN galaxies reveal a correlation with
metallicity, towards high PC2 projections having a higher [Z/H], and
SF may just show a mild correlation with age, with the low PC2 distribution
hinting at a tail towards older ages.
PC3 behaves similarly to PC2 in Q and SF galaxies and 
anti-correlates with metallicity in AGN spectra.
Appendix~\ref{Sec:ApSpecRed} shows the confidence levels
as in Fig.~\ref{fig:MCMC} for the red spectral interval.
The trends are strongly consistent for the projections on the
first principal component, reflecting the substantial covariance
across a wide range of spectra wavelengths \citep[as illustrated in][]{InfoPop}.
The comparison is more subtle with PC2 and PC3. For instance, PC2 keeps a
positive correlation in AGN with respect to metallicity.

\section{Discussion and Conclusions}
\label{Sec:Disc}

The focus of this paper is to explore galaxy spectra by
relying exclusively on the data to extract information about the
underlying stellar populations. This approach is complementary to
standard methods based on population synthesis modelling 
\citep[see, e.g.,][]{Starlight:05,Vespa:07,ppxf:12,Firefly:17,Robotham:20}. 
The ``front end'' of these models depend on a 
small set of parameters that allow a relatively simple 
fit of photometric and spectroscopic observations. These parameters
relate to the age, metallicity, chemical
enrichment, star formation, or stellar mass.
The model-based studies are limited due to the high
entanglement and complexity in the  definition of parameter space,
with a number of strong degeneracies, such as the one between 
age, metallicity and dust \citep{W:94,FCS:99}. In
this study, as explained in Sec.~\ref{Sec:PCA} , we adopt a
multivariate analysis method, principal component analysis, as a blind
source separation technique to decipher the information encoded in the
optical spectra of $\sim$70000 galaxies from the Legacy part of
SDSS. We only use population synthesis models a posteriori, to gain 
physical insight of the components retrieved by the statistical
analysis.

As the analysis is purely dependent on the input data, we need to
make sure the spectra have a high enough signal to noise ratio, and,
most importantly, we remove the pseudo-continuum and avoid spectral
intervals dominated by emission lines, so that the contribution from
dust and diffuse gas is minimised. At the same time, we need to focus
the data as much as possible, to avoid being overwhelmed by well-known
trends, such as the age-mass and metallicity-mass
relations \citep[e.g.][]{Gallazzi:05}, or by
the effect of stellar velocity dispersion on the effective resolution
of the spectra. Therefore, we also restrict the sample to relatively
low velocity dispersion ($\sigma\in[100,150]$\,km/s). Moreover, 
we split the sample into three subsets, according to the nebular
emission properties into quiescent, AGN and star forming, noting, though,
that the input data does not include the flux from emission lines, so
that the observed variance only depends on the absorption lines originating
in the atmospheres of the stellar populations. Concerning variations 
in the stellar initial mass function (IMF), such as the observed bottom-heavy phase in
the cores of massive early-types \cite[e.g.,][]{vDC:10,IMF:13,FLB:19} or
a top-heavy phase in star-forming galaxies \citep[e.g.,][]{Lee:09,Madusha:11},
we emphasize that our sample is representative of a general, relatively
low-mass population, for which a standard IMF gives a good representation.
Note also that the signatures typically found in absorption line spectra
from IMF variations stay at the sub-percent level \citep[see, e.g.,][]{FLB:13}.

\begin{figure*}
\includegraphics[width=85mm]{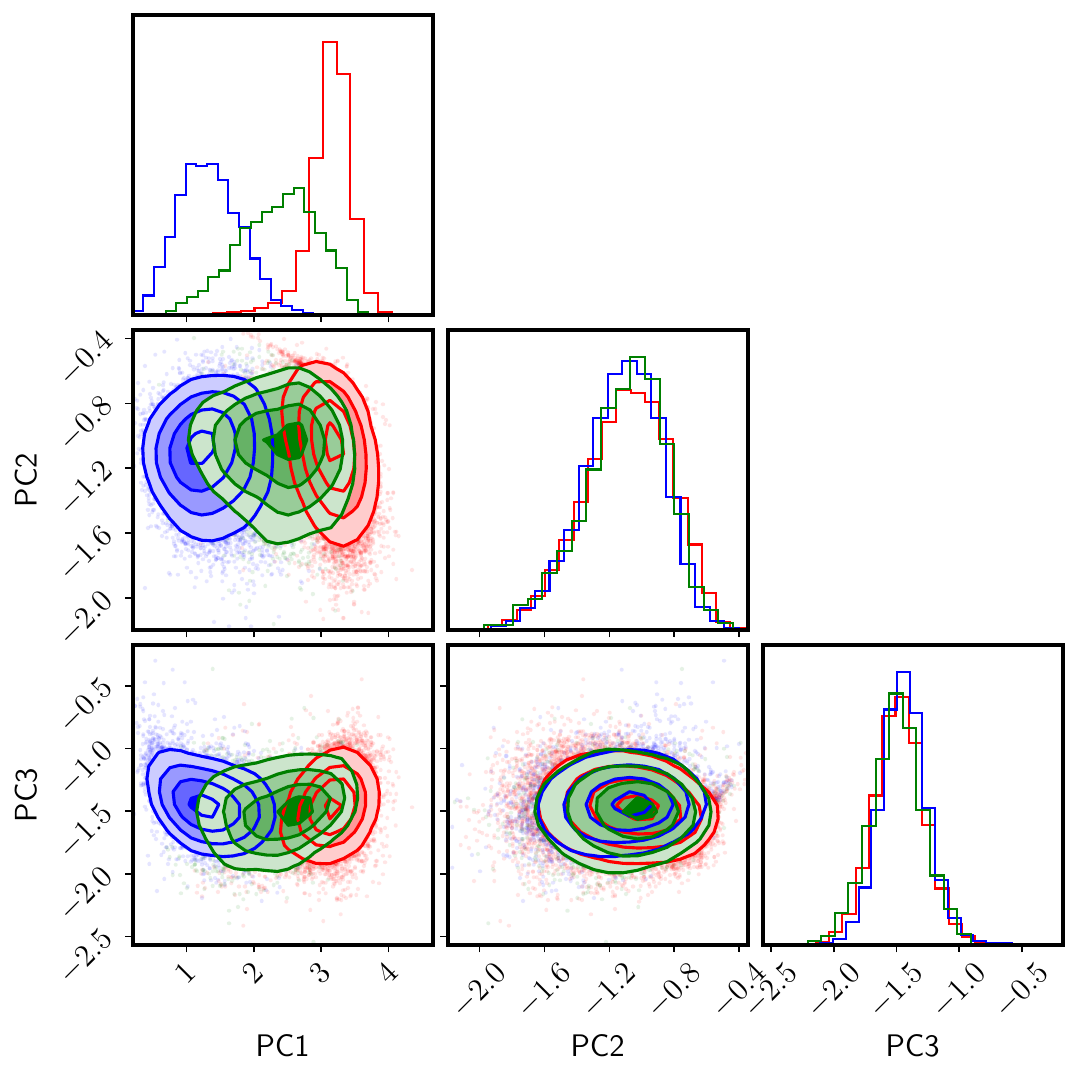}
\includegraphics[width=85mm]{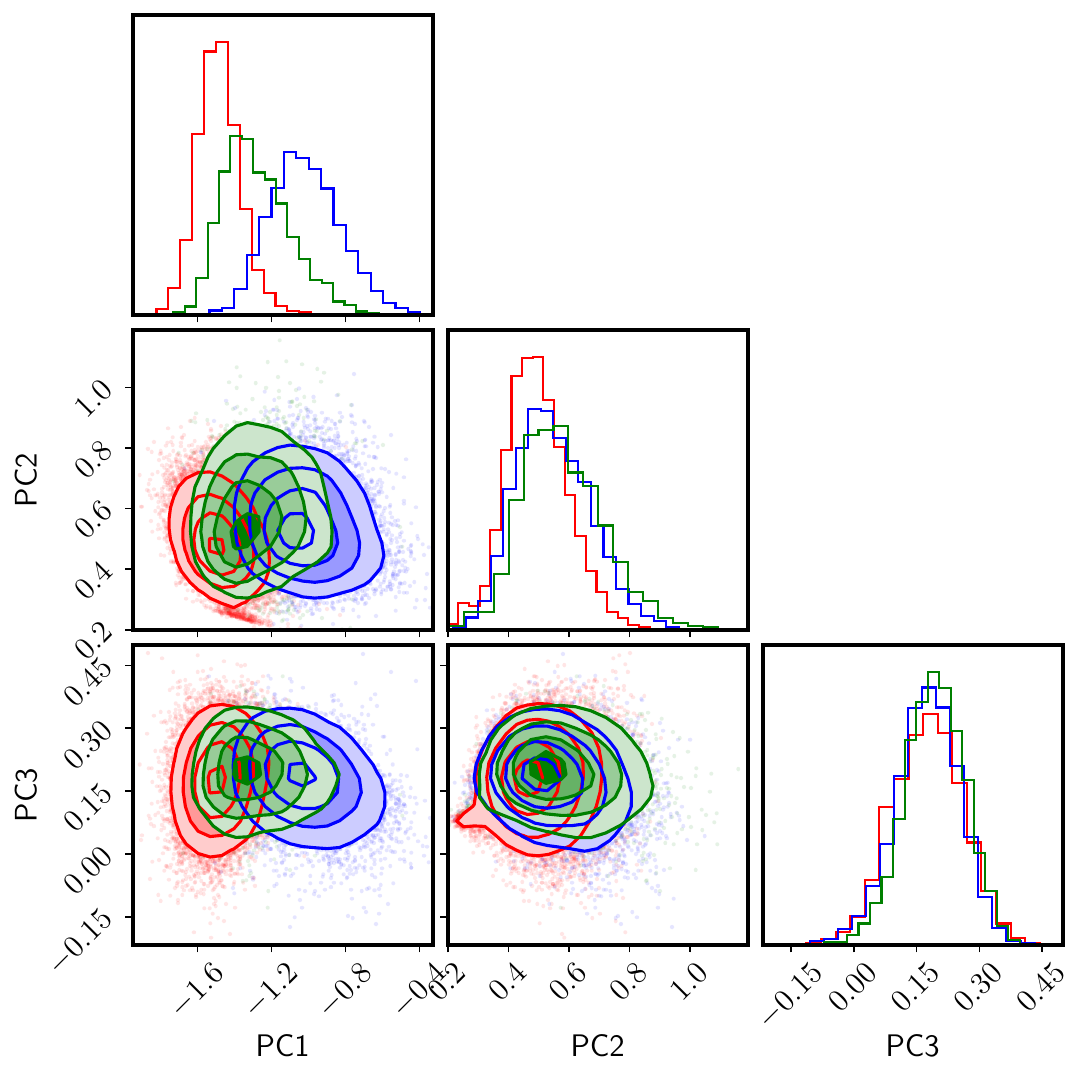}
\caption{Distribution of principal component projections as in
  Fig.~\ref{fig:corner}, treating the whole sample as one set,
  i.e. diagonalising a single covariance matrix, and separating
  the spectra according to their nebular emissio properties
  a posteriori. The left (right) panels show the result for the
blue (red) interval. The contours engulf 25, 50, 75 and 90\% of each subsample.}
\label{fig:CornerAll}
\end{figure*}

Previous work applying PCA to galaxy spectra finds a high
compressibility, showing, for instance, that over 90\% of the total
variance is encoded in a single principal component
\citep{Rogers:07}. However, our data (Fig.~\ref{fig:scree}) show a
more spread distribution of the weights, which reflects the high level
of correlatedness of the pseudo-continuum, that is removed in our
analysis. In our data, the first component holds up to $\sim$10\% of
total variance, where Q galaxies have the lowest variance. This is a
reflection of the fact that quiescent galaxies have a more homogeneous
distribution of (older) populations, but it is noteworthy to compare
SF and AGN subsets, with the latter consistently keeping a higher
fraction of variance in both spectral windows.

By projecting the observed spectra on to the first three principal
components, we reduce the dimensionality of the problem, describing
each galaxy by a set of three numbers that keep the highest amount of
variance. Our aim is to understand how these components relate to
physical properties. The trends shown in Figs.~\ref{fig:Corr1} and
\ref{fig:Corr2} show that the first component is strongly correlated
with traditional spectral indicators (4000\AA\ break strength, Balmer
lines and Mg and Fe absorption). However, in the higher order components,
the trend
is less trivial. For instance, PC3 cleanly separates AGN from the
other galaxies in both spectral intervals (see Fig.~\ref{fig:corner}),
with milder trends with respect to the line strengths, except for
H$\delta_A$ in SF galaxies for the blue interval and Mgb in AGN+SF, as
well as $\langle$Fe$\rangle$ in Q for the red interval. The
distribution of PC2 projections feature the largest separation between
the three subsets Q/AGN/SF in both spectral intervals, and have mixed
behaviour regarding the trends with the line strengths. Turning to the
other parameters, a trend is found with velocity dispersion, which can
be caused by a combination of a physical correlation, as $\sigma$
traces the gravitational potential of the galaxy and hence its
formation history, but there is also an instrumental effect, as
$\sigma$ will affect the effective resolution of the spectra. We
cannot ascribe any of these parameters as major drivers of evolution,
similarly to traditional work \citep{Ferreras:19}

Projecting synthetic spectra of simple stellar populations, as shown in 
Figs.~\ref{fig:SSPtracks} and \ref{fig:animation2}, confirms that PC1 in both spectral
intervals strongly correlates with stellar age.  It is worth
mentioning that PC2 -- that clearly separates the three types -- is
more entangled with SSP parameters, with the red interval PC2 showing
monotonic behaviour with age, but the blue interval PC2 featuring a
more complex evolution, with a characteristic kink at young ages.  PC3
is even less predictable from an SSP based analysis.  The SSP tracks
also provide an excellent illustration of
the age-metallicity degeneracy. Population synthesis models have
already shown this entanglement \citep[e.g.][]{W:94,FCS:99}, but the
variance analysis gives a more robust visualization, so that changes in
either SSP-equivalent age or metallicity evolve very much in the same
direction, even up to the third principal component. This is a sobering
reminder of the difficulty in extracting information from galaxy
spectra.

We also perfomed spectral fits to stacked data based on the highest
and lowest percentiles of the distribution of PC1, PC2 and PC3
(Fig.~\ref{fig:MCMC}). PC1 consistently maps SSP-equivalent age in the
three subsets, and PC2 and PC3 show a mixed dependence with respect to the
SF/AGN/Q nature of the galaxy. Q galaxies appear not to feature any
substantial difference with respect to high/low values of PC2 and PC3. 
AGN galaxies have a significant trend in PC2 and PC3 with metallicity,
and the SF subset show high overlap of the parameter distribution, but with a
hint of a tail in the probability distribution towards older ages in
galaxies with the lowest projections. Our PCA
decomposition fully depends on absorption lines (neither continuum
nor emission lines are included), and aligns with the hypothesis of 
AGN statistically representing a transitioning phase between star
formation and quiescence \citep{Schawinski:07}. This work, however,
cannot reject the hypothesis of a quenching channel purely driven by star
formation. We emphasize that
these three components are based on a variance analysis of
observational data. However, the idiosyncrasy of PCA, for instance the
enforced orthogonality of the eigenvectors, may result in unphysical
components. The way forward, reserved to future work, will involve
comparisons with realistic models of galaxy formation and evolution
(Sharbaf et al., in preparation).

The inquisitive reader might ponder about the way we separated the
sample in three sets, producing three different covariances. The
motivation is to maximise the ability of PCA to retrieve a meaningful
signal. By putting all the galaxies together, we produce a more mixed
population where the first few components will be most sensitive to
the mixture. However, we include another test by diagonalising a single
covariance, i.e. putting together the whole sample regardless of the
SF/AGN/Q nature. In addition, we do not cull the sample according to
eq.~\ref{eq:4sigma}, and show the projections in Fig.~\ref{fig:CornerAll},
where the colour-coded separation is perfomed {\sl after} the
diagonalisation, i.e. there is a single set of eigenvectors.
Note PC1 consistently finds a trend between the three
classes. However, the mixture of spectra results in no significant
differences in the higher order components, justifying our original approach,
and serving as a cautionary advice regarding the definition of the
input data even in the most sophisticated Machine Learning algorithms.

\begin{figure}
  \includegraphics[width=85mm]{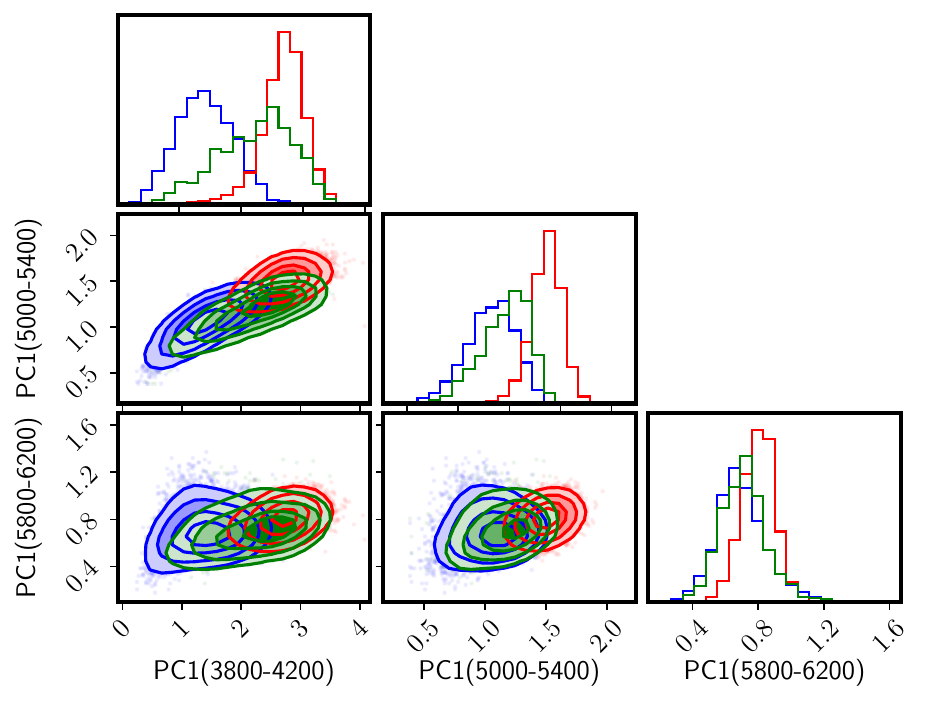}
  \caption{Projections of the first principal component, analogously
    to Fig.~\ref{fig:corner} in three different spectral windows, as
    labelled (the wavelength range is shown in \AA).  This figure
    illustrates the higher information content (understood as
    variance) in the bluer interval, that roughly maps the
    4000\AA\ break. The contours engulf 25, 50, 75 and 90\% of each
    subsample.}
  \label{fig:corner3BPs}
\end{figure}

Finally, we also explored the choice of spectral window. Throughout this
paper we consider two intervals, where the trends found are consistent
although not identical. Fig.~\ref{fig:corner3BPs} shows the projection
of the data on to the first principal component in three different
spectral windows: the blue and red ones defined thus far, along with a
redder one, defined in the [5800,6200]\AA\ wavelength range. The results
are consistent, with a strong correlation between the PC1 projections in
the blue and red intervals. Note the separation between the three subsets, thus
regarding their different evolutionary state, is maximised in the bluer
interval, in line with entropy-based studies that show a higher information
content towards bluer wavelengths \citep{InfoPop}. At the reddest interval, the
separation is less clear, as the absorption lines vary much less, and
thus would require a higher S/N ratio to produce meaningful results.

We conclude in a more speculative tone: All the tests investigated in
this study through PCA suggest that the structure of galaxies may be
controlled by a single (or a few) parameters, as we can see through
this result that only one component  shows a correlation with age, and 
plays a main role as an evolutionary trend. Although the evolution of
galaxies based on cold dark matter suggests that the properties of a
galaxy should be controlled by a large range of physical parameters
\citep{Baugh:06,Dalcanton:97, MMW:98}, our
study is reminiscent of \citep{Disney:08}, who found that a
single parameter could ``explain'' most of the fundamental
scaling trends of galaxies.

\section*{Acknowledgements}
IF and ZS acknowledge support from the Spanish Research Agency of the
Ministry of Science and Innovation (AEI-MICINN) under grant 
PID2019-104788GB-I00. OL acknowledges STFC Consolidated
Grant ST/R000476/1 and a Visiting Fellowship at All Souls College,
Oxford. Funding for SDSS-III has been provided by
the Alfred P. Sloan Foundation, the Participating Institutions, the
National Science Foundation, and the U.S. Department of Energy Office
of Science. The SDSS-III web site is http://www.sdss3.org/.

\section*{Data availability}
This work has been fully based on publicly available data: 
galaxy spectra were retrieved from the SDSS DR16 archive
(https://www.sdss.org/dr16/) and stellar population
synthesis models can be obtained from the respective authors.



\bibliographystyle{mnras}
\bibliography{PCA23_SFL} 

\begin{thebibliography}{}
\makeatletter
\relax
\def\mn@urlcharsother{\let\do\@makeother \do\$\do\&\do\#\do\^\do\_\do\%\do\~}
\def\mn@doi{\begingroup\mn@urlcharsother \@ifnextchar [ {\mn@doi@}
  {\mn@doi@[]}}
\def\mn@doi@[#1]#2{\def\@tempa{#1}\ifx\@tempa\@empty \href
  {http://dx.doi.org/#2} {doi:#2}\else \href {http://dx.doi.org/#2} {#1}\fi
  \endgroup}
\def\mn@eprint#1#2{\mn@eprint@#1:#2::\@nil}
\def\mn@eprint@arXiv#1{\href {http://arxiv.org/abs/#1} {{\tt arXiv:#1}}}
\def\mn@eprint@dblp#1{\href {http://dblp.uni-trier.de/rec/bibtex/#1.xml}
  {dblp:#1}}
\def\mn@eprint@#1:#2:#3:#4\@nil{\def\@tempa {#1}\def\@tempb {#2}\def\@tempc
  {#3}\ifx \@tempc \@empty \let \@tempc \@tempb \let \@tempb \@tempa \fi \ifx
  \@tempb \@empty \def\@tempb {arXiv}\fi \@ifundefined
  {mn@eprint@\@tempb}{\@tempb:\@tempc}{\expandafter \expandafter \csname
  mn@eprint@\@tempb\endcsname \expandafter{\@tempc}}}

\bibitem[\protect\citeauthoryear{{Ahumada} et~al.}{{Ahumada}
  et~al.}{2020}]{DR16}
{Ahumada} et~al., 2020, \mn@doi [\apjs] {10.3847/1538-4365/ab929e}, \href
  {https://ui.adsabs.harvard.edu/abs/2020ApJS..249....3A} {249, 3}

\bibitem[\protect\citeauthoryear{{Angthopo}, {Ferreras}  \& {Silk}}{{Angthopo}
  et~al.}{2019}]{Angthopo:19}
{Angthopo} J.,  {Ferreras} I.,   {Silk} J.,  2019, \mn@doi [\mnras]
  {10.1093/mnrasl/slz106}, \href
  {https://ui.adsabs.harvard.edu/abs/2019MNRAS.488L..99A} {488, L99}

\bibitem[\protect\citeauthoryear{{Angthopo}, {Ferreras}  \& {Silk}}{{Angthopo}
  et~al.}{2020}]{Angthopo:20}
{Angthopo} J.,  {Ferreras} I.,   {Silk} J.,  2020, \mn@doi [\mnras]
  {10.1093/mnras/staa1276}, \href
  {https://ui.adsabs.harvard.edu/abs/2020MNRAS.495.2720A} {495, 2720}

\bibitem[\protect\citeauthoryear{{Baldry}, {Glazebrook}, {Brinkmann},
  {Ivezi{\'c}}, {Lupton}, {Nichol}  \& {Szalay}}{{Baldry}
  et~al.}{2004}]{Baldry:04}
{Baldry} I.~K.,  {Glazebrook} K.,  {Brinkmann} J.,  {Ivezi{\'c}} {\v{Z}}.,
  {Lupton} R.~H.,  {Nichol} R.~C.,   {Szalay} A.~S.,  2004, \mn@doi [\apj]
  {10.1086/380092}, \href
  {https://ui.adsabs.harvard.edu/abs/2004ApJ...600..681B} {600, 681}

\bibitem[\protect\citeauthoryear{{Baldwin}, {Phillips}  \&
  {Terlevich}}{{Baldwin} et~al.}{1981}]{BPT}
{Baldwin} J.~A.,  {Phillips} M.~M.,   {Terlevich} R.,  1981, \mn@doi [\pasp]
  {10.1086/130766}, \href
  {https://ui.adsabs.harvard.edu/abs/1981PASP...93....5B} {93, 5}

\bibitem[\protect\citeauthoryear{{Ball}, {Loveday}, {Fukugita}, {Nakamura},
  {Okamura}, {Brinkmann}  \& {Brunner}}{{Ball} et~al.}{2004}]{Ball:04}
{Ball} N.~M.,  {Loveday} J.,  {Fukugita} M.,  {Nakamura} O.,  {Okamura} S.,
  {Brinkmann} J.,   {Brunner} R.~J.,  2004, \mn@doi [\mnras]
  {10.1111/j.1365-2966.2004.07429.x}, \href
  {https://ui.adsabs.harvard.edu/abs/2004MNRAS.348.1038B} {348, 1038}

\bibitem[\protect\citeauthoryear{{Balogh}, {Morris}, {Yee}, {Carlberg}  \&
  {Ellingson}}{{Balogh} et~al.}{1999}]{Balogh:99}
{Balogh} M.~L.,  {Morris} S.~L.,  {Yee} H.~K.~C.,  {Carlberg} R.~G.,
  {Ellingson} E.,  1999, \mn@doi [\apj] {10.1086/308056}, \href
  {https://ui.adsabs.harvard.edu/abs/1999ApJ...527...54B} {527, 54}

\bibitem[\protect\citeauthoryear{{Baugh}}{{Baugh}}{2006}]{Baugh:06}
{Baugh} C.~M.,  2006, \mn@doi [Reports on Progress in Physics]
  {10.1088/0034-4885/69/12/R02}, \href
  {https://ui.adsabs.harvard.edu/abs/2006RPPh...69.3101B} {69, 3101}

\bibitem[\protect\citeauthoryear{{Bernardi} et~al.}{{Bernardi}
  et~al.}{2003}]{Bernardi:03}
{Bernardi} et~al., 2003, \mn@doi [\aj] {10.1086/374256}, \href
  {https://ui.adsabs.harvard.edu/abs/2003AJ....125.1849B} {125, 1849}

\bibitem[\protect\citeauthoryear{{Brinchmann}, {Charlot}, {White}, {Tremonti},
  {Kauffmann}, {Heckman}  \& {Brinkmann}}{{Brinchmann}
  et~al.}{2004}]{Brinchmann:04}
{Brinchmann} J.,  {Charlot} S.,  {White} S.~D.~M.,  {Tremonti} C.,  {Kauffmann}
  G.,  {Heckman} T.,   {Brinkmann} J.,  2004, \mn@doi [\mnras]
  {10.1111/j.1365-2966.2004.07881.x}, \href
  {https://ui.adsabs.harvard.edu/abs/2004MNRAS.351.1151B} {351, 1151}

\bibitem[\protect\citeauthoryear{{Bruzual} \& {Charlot}}{{Bruzual} \&
  {Charlot}}{2003}]{BC:03}
{Bruzual} G.,  {Charlot} S.,  2003, \mn@doi [\mnras]
  {10.1046/j.1365-8711.2003.06897.x}, \href
  {https://ui.adsabs.harvard.edu/abs/2003MNRAS.344.1000B} {344, 1000}

\bibitem[\protect\citeauthoryear{{Bruzual A.} \& {Charlot}}{{Bruzual A.} \&
  {Charlot}}{1993}]{Bruzual:93}
{Bruzual A.} G.,  {Charlot} S.,  1993, \mn@doi [\apj] {10.1086/172385}, \href
  {https://ui.adsabs.harvard.edu/abs/1993ApJ...405..538B} {405, 538}

\bibitem[\protect\citeauthoryear{{Cappellari}}{{Cappellari}}{2012}]{ppxf:12}
{Cappellari} M.,  2012, {pPXF: Penalized Pixel-Fitting stellar kinematics
  extraction}, Astrophysics Source Code Library, record ascl:1210.002
  (\mn@eprint {ascl} {1210.002})

\bibitem[\protect\citeauthoryear{{Cid Fernandes}, {Mateus}, {Sodr{\'e}},
  {Stasi{\'n}ska}  \& {Gomes}}{{Cid Fernandes} et~al.}{2005}]{Starlight:05}
{Cid Fernandes} R.,  {Mateus} A.,  {Sodr{\'e}} L.,  {Stasi{\'n}ska} G.,
  {Gomes} J.~M.,  2005, \mn@doi [\mnras] {10.1111/j.1365-2966.2005.08752.x},
  \href {https://ui.adsabs.harvard.edu/abs/2005MNRAS.358..363C} {358, 363}

\bibitem[\protect\citeauthoryear{{Cid Fernandes}, {Stasi{\'n}ska}, {Mateus}  \&
  {Vale Asari}}{{Cid Fernandes} et~al.}{2011}]{CF:11}
{Cid Fernandes} R.,  {Stasi{\'n}ska} G.,  {Mateus} A.,   {Vale Asari} N.,
  2011, \mn@doi [\mnras] {10.1111/j.1365-2966.2011.18244.x}, \href
  {https://ui.adsabs.harvard.edu/abs/2011MNRAS.413.1687C} {413, 1687}

\bibitem[\protect\citeauthoryear{{Conroy}}{{Conroy}}{2013}]{Conroy:13}
{Conroy} C.,  2013, \mn@doi [ARA\& A] {10.1146/annurev-astro-082812-141017},
  \href {https://ui.adsabs.harvard.edu/abs/2013ARA&A..51..393C} {51, 393}

\bibitem[\protect\citeauthoryear{{Conroy} \& {Gunn}}{{Conroy} \&
  {Gunn}}{2010}]{Conroy:10}
{Conroy} C.,  {Gunn} J.~E.,  2010, \mn@doi [\apj]
  {10.1088/0004-637X/712/2/833}, \href
  {https://ui.adsabs.harvard.edu/abs/2010ApJ...712..833C} {712, 833}

\bibitem[\protect\citeauthoryear{{Dalcanton}, {Spergel}  \&
  {Summers}}{{Dalcanton} et~al.}{1997}]{Dalcanton:97}
{Dalcanton} J.~J.,  {Spergel} D.~N.,   {Summers} F.~J.,  1997, \mn@doi [\apj]
  {10.1086/304182}, \href
  {https://ui.adsabs.harvard.edu/abs/1997ApJ...482..659D} {482, 659}

\bibitem[\protect\citeauthoryear{{Disney}, {Romano}, {Garcia-Appadoo}, {West},
  {Dalcanton}  \& {Cortese}}{{Disney} et~al.}{2008}]{Disney:08}
{Disney} M.~J.,  {Romano} J.~D.,  {Garcia-Appadoo} D.~A.,  {West} A.~A.,
  {Dalcanton} J.~J.,   {Cortese} L.,  2008, \mn@doi [\nat]
  {10.1038/nature07366}, \href
  {https://ui.adsabs.harvard.edu/abs/2008Natur.455.1082D} {455, 1082}

\bibitem[\protect\citeauthoryear{{Ferreras} \& {Silk}}{{Ferreras} \&
  {Silk}}{2000}]{FS:00}
{Ferreras} I.,  {Silk} J.,  2000, \mn@doi [\apjl] {10.1086/312898}, \href
  {https://ui.adsabs.harvard.edu/abs/2000ApJ...541L..37F} {541, L37}

\bibitem[\protect\citeauthoryear{{Ferreras}, {Charlot}  \& {Silk}}{{Ferreras}
  et~al.}{1999}]{FCS:99}
{Ferreras} I.,  {Charlot} S.,   {Silk} J.,  1999, \mn@doi [\apj]
  {10.1086/307513}, \href
  {https://ui.adsabs.harvard.edu/abs/1999ApJ...521...81F} {521, 81}

\bibitem[\protect\citeauthoryear{{Ferreras}, {Pasquali}, {de Carvalho}, {de la
  Rosa}  \& {Lahav}}{{Ferreras} et~al.}{2006}]{Ferreras:06}
{Ferreras} I.,  {Pasquali} A.,  {de Carvalho} R.~R.,  {de la Rosa} I.~G.,
  {Lahav} O.,  2006, \mn@doi [\mnras] {10.1111/j.1365-2966.2006.10509.x}, \href
  {https://ui.adsabs.harvard.edu/abs/2006MNRAS.370..828F} {370, 828}

\bibitem[\protect\citeauthoryear{{Ferreras}, {La Barbera}, {de La Rosa},
  {Vazdekis}, {de Carvalho}, {Falcon-Barroso}  \& {Ricciardelli}}{{Ferreras}
  et~al.}{2013}]{IMF:13}
{Ferreras} I.,  {La Barbera} F.,  {de La Rosa} I.~G.,  {Vazdekis} A.,  {de
  Carvalho} R.~R.,  {Falcon-Barroso} J.,   {Ricciardelli} E.,  2013, \mn@doi
  [\mnras] {10.1093/mnrasl/sls014}, \href
  {https://ui.adsabs.harvard.edu/abs/2013MNRAS.429L..15F} {429, L15}

\bibitem[\protect\citeauthoryear{{Ferreras}, {Hopkins}, {Lagos}, {Sansom},
  {Scott}, {Croom}  \& {Brough}}{{Ferreras} et~al.}{2019a}]{SDSSCPs}
{Ferreras} I.,  {Hopkins} A.~M.,  {Lagos} C.,  {Sansom} A.~E.,  {Scott} N.,
  {Croom} S.,   {Brough} S.,  2019a, \mn@doi [\mnras] {10.1093/mnras/stz1286},
  \href {https://ui.adsabs.harvard.edu/abs/2019MNRAS.487..435F} {487, 435}

\bibitem[\protect\citeauthoryear{{Ferreras} et~al.,}{{Ferreras}
  et~al.}{2019b}]{Ferreras:19}
{Ferreras} I.,  et~al., 2019b, \mn@doi [\mnras] {10.1093/mnras/stz2095}, \href
  {https://ui.adsabs.harvard.edu/abs/2019MNRAS.489..608F} {489, 608}

\bibitem[\protect\citeauthoryear{{Ferreras}, {Lahav}, {Somerville}  \&
  {Silk}}{{Ferreras} et~al.}{2023}]{InfoPop}
{Ferreras} I.,  {Lahav} O.,  {Somerville} R.~S.,   {Silk} J.,  2023, \mn@doi
  [RAS Techniques and Instruments] {10.1093/rasti/rzad004}, \href
  {https://ui.adsabs.harvard.edu/abs/2023RASTI...2...78F} {2, 78}

\bibitem[\protect\citeauthoryear{{Fioc} \& {Rocca-Volmerange}}{{Fioc} \&
  {Rocca-Volmerange}}{1997}]{Fioc:97}
{Fioc} M.,  {Rocca-Volmerange} B.,  1997, \mn@doi [\aap]
  {10.48550/arXiv.astro-ph/9707017}, \href
  {https://ui.adsabs.harvard.edu/abs/1997A&A...326..950F} {326, 950}

\bibitem[\protect\citeauthoryear{Folkes, Lahav  \& Maddox}{Folkes
  et~al.}{1996}]{folkes1996}
Folkes S.,  Lahav O.,   Maddox S.,  1996, \mnras, 283, 651

\bibitem[\protect\citeauthoryear{Foreman-Mackey}{Foreman-Mackey}{2016}]{corner}
Foreman-Mackey D.,  2016, \mn@doi [The Journal of Open Source Software]
  {10.21105/joss.00024}, 1, 24

\bibitem[\protect\citeauthoryear{{Foreman-Mackey}, {Hogg}, {Lang}  \&
  {Goodman}}{{Foreman-Mackey} et~al.}{2013}]{emcee}
{Foreman-Mackey} D.,  {Hogg} D.~W.,  {Lang} D.,   {Goodman} J.,  2013, \mn@doi
  [PASP] {10.1086/670067}, \href
  {https://ui.adsabs.harvard.edu/abs/2013PASP..125..306F} {125, 306}

\bibitem[\protect\citeauthoryear{{Gallazzi}, {Charlot}, {Brinchmann}, {White}
  \& {Tremonti}}{{Gallazzi} et~al.}{2005}]{Gallazzi:05}
{Gallazzi} A.,  {Charlot} S.,  {Brinchmann} J.,  {White} S. D.~M.,   {Tremonti}
  C.~A.,  2005, \mn@doi [\mnras] {10.1111/j.1365-2966.2005.09321.x}, \href
  {https://ui.adsabs.harvard.edu/abs/2005MNRAS.362...41G} {362, 41}

\bibitem[\protect\citeauthoryear{{Gallazzi}, {Bell}, {Zibetti}, {Brinchmann}
  \& {Kelson}}{{Gallazzi} et~al.}{2014}]{Gallazzi:14}
{Gallazzi} A.,  {Bell} E.~F.,  {Zibetti} S.,  {Brinchmann} J.,   {Kelson}
  D.~D.,  2014, \mn@doi [\apj] {10.1088/0004-637X/788/1/72}, \href
  {https://ui.adsabs.harvard.edu/abs/2014ApJ...788...72G} {788, 72}

\bibitem[\protect\citeauthoryear{{Ge}, {Yan}, {Cappellari}, {Mao}, {Li}  \&
  {Lu}}{{Ge} et~al.}{2018}]{Ge:18}
{Ge} J.,  {Yan} R.,  {Cappellari} M.,  {Mao} S.,  {Li} H.,   {Lu} Y.,  2018,
  \mn@doi [\mnras] {10.1093/mnras/sty1245}, \href
  {https://ui.adsabs.harvard.edu/abs/2018MNRAS.478.2633G} {478, 2633}

\bibitem[\protect\citeauthoryear{{Ge}, {Mao}, {Lu}, {Cappellari}  \&
  {Yan}}{{Ge} et~al.}{2019}]{Ge:19}
{Ge} J.,  {Mao} S.,  {Lu} Y.,  {Cappellari} M.,   {Yan} R.,  2019, \mn@doi
  [\mnras] {10.1093/mnras/stz418}, \href
  {https://ui.adsabs.harvard.edu/abs/2019MNRAS.485.1675G} {485, 1675}

\bibitem[\protect\citeauthoryear{{Gunawardhana} et~al.,}{{Gunawardhana}
  et~al.}{2011}]{Madusha:11}
{Gunawardhana} M.~L.~P.,  et~al., 2011, \mn@doi [\mnras]
  {10.1111/j.1365-2966.2011.18800.x}, \href
  {https://ui.adsabs.harvard.edu/abs/2011MNRAS.415.1647G} {415, 1647}

\bibitem[\protect\citeauthoryear{{Hart}}{{Hart}}{2019}]{Hart:19}
{Hart} M.,  2019, \mn@doi [\aj] {10.3847/1538-3881/ab1a35}, \href
  {https://ui.adsabs.harvard.edu/abs/2019AJ....157..221H} {157, 221}

\bibitem[\protect\citeauthoryear{{Hawkins}, {Jofr{\'e}}, {Gilmore}  \&
  {Masseron}}{{Hawkins} et~al.}{2014}]{Hawkins:14}
{Hawkins} K.,  {Jofr{\'e}} P.,  {Gilmore} G.,   {Masseron} T.,  2014, \mn@doi
  [\mnras] {10.1093/mnras/stu1910}, \href
  {https://ui.adsabs.harvard.edu/abs/2014MNRAS.445.2575H} {445, 2575}

\bibitem[\protect\citeauthoryear{{Huertas-Company} \&
  {Lanusse}}{{Huertas-Company} \& {Lanusse}}{2023}]{MHC:23}
{Huertas-Company} M.,  {Lanusse} F.,  2023, \mn@doi [\pasa]
  {10.1017/pasa.2022.55}, \href
  {https://ui.adsabs.harvard.edu/abs/2023PASA...40....1H} {40, e001}

\bibitem[\protect\citeauthoryear{{Kaban}, {Nolan}  \& {Raychaudhury}}{{Kaban}
  et~al.}{2005}]{Kaban:05}
{Kaban} A.,  {Nolan} L.~A.,   {Raychaudhury} S.,  2005, in Proceedings of the
  Fifth SIAM International Conference on Data Mining Editors: Hillol Kargupta.
  p.~183 (\mn@eprint {arXiv} {astro-ph/0505059}),
  \mn@doi{10.48550/arXiv.astro-ph/0505059}

\bibitem[\protect\citeauthoryear{{Karhunen}}{{Karhunen}}{1947}]{Karhunen:47}
{Karhunen} 1947, Ann. Acad. Sci. Fennicae, 37, 1

\bibitem[\protect\citeauthoryear{{Kaviraj} et~al.,}{{Kaviraj}
  et~al.}{2007}]{Kaviraj:07}
{Kaviraj} S.,  et~al., 2007, \mn@doi [\apjs] {10.1086/516633}, \href
  {https://ui.adsabs.harvard.edu/abs/2007ApJS..173..619K} {173, 619}

\bibitem[\protect\citeauthoryear{{Kroupa}}{{Kroupa}}{2001}]{KroupaU:01}
{Kroupa} P.,  2001, \mn@doi [\mnras] {10.1046/j.1365-8711.2001.04022.x}, \href
  {https://ui.adsabs.harvard.edu/abs/2001MNRAS.322..231K} {322, 231}

\bibitem[\protect\citeauthoryear{{Kuntschner} et~al.,}{{Kuntschner}
  et~al.}{2010}]{Kuntschner:10}
{Kuntschner} H.,  et~al., 2010, \mn@doi [\mnras]
  {10.1111/j.1365-2966.2010.17161.x}, \href
  {https://ui.adsabs.harvard.edu/abs/2010MNRAS.408...97K} {408, 97}

\bibitem[\protect\citeauthoryear{{La Barbera}, {Ferreras}, {Vazdekis}, {de la
  Rosa}, {de Carvalho}, {Trevisan}, {Falc{\'o}n-Barroso}  \&
  {Ricciardelli}}{{La Barbera} et~al.}{2013}]{FLB:13}
{La Barbera} F.,  {Ferreras} I.,  {Vazdekis} A.,  {de la Rosa} I.~G.,  {de
  Carvalho} R.~R.,  {Trevisan} M.,  {Falc{\'o}n-Barroso} J.,   {Ricciardelli}
  E.,  2013, \mn@doi [\mnras] {10.1093/mnras/stt943}, \href
  {https://ui.adsabs.harvard.edu/abs/2013MNRAS.433.3017L} {433, 3017}

\bibitem[\protect\citeauthoryear{{La Barbera}, {Pasquali}, {Ferreras},
  {Gallazzi}, {de Carvalho}  \& {de la Rosa}}{{La Barbera}
  et~al.}{2014}]{FLB:14}
{La Barbera} F.,  {Pasquali} A.,  {Ferreras} I.,  {Gallazzi} A.,  {de Carvalho}
  R.~R.,   {de la Rosa} I.~G.,  2014, \mn@doi [\mnras] {10.1093/mnras/stu1626},
  \href {https://ui.adsabs.harvard.edu/abs/2014MNRAS.445.1977L} {445, 1977}

\bibitem[\protect\citeauthoryear{{La Barbera} et~al.,}{{La Barbera}
  et~al.}{2019}]{FLB:19}
{La Barbera} F.,  et~al., 2019, \mn@doi [\mnras] {10.1093/mnras/stz2192}, \href
  {https://ui.adsabs.harvard.edu/abs/2019MNRAS.489.4090L} {489, 4090}

\bibitem[\protect\citeauthoryear{{Lee} et~al.,}{{Lee} et~al.}{2009}]{Lee:09}
{Lee} J.~C.,  et~al., 2009, \mn@doi [\apj] {10.1088/0004-637X/706/1/599}, \href
  {https://ui.adsabs.harvard.edu/abs/2009ApJ...706..599L} {706, 599}

\bibitem[\protect\citeauthoryear{{Lovell}, {Acquaviva}, {Thomas}, {Iyer},
  {Gawiser}  \& {Wilkins}}{{Lovell} et~al.}{2019}]{Lovell:19}
{Lovell} C.~C.,  {Acquaviva} V.,  {Thomas} P.~A.,  {Iyer} K.~G.,  {Gawiser} E.,
    {Wilkins} S.~M.,  2019, \mn@doi [\mnras] {10.1093/mnras/stz2851}, \href
  {https://ui.adsabs.harvard.edu/abs/2019MNRAS.490.5503L} {490, 5503}

\bibitem[\protect\citeauthoryear{{Lu}, {Zhou}, {Wang}, {Wang}, {Dong}, {Zhuang}
   \& {Li}}{{Lu} et~al.}{2006}]{Lu:06}
{Lu} H.,  {Zhou} H.,  {Wang} J.,  {Wang} T.,  {Dong} X.,  {Zhuang} Z.,   {Li}
  C.,  2006, \mn@doi [\aj] {10.1086/498711}, \href
  {https://ui.adsabs.harvard.edu/abs/2006AJ....131..790L} {131, 790}

\bibitem[\protect\citeauthoryear{Madgwick et~al.,}{Madgwick
  et~al.}{2003}]{Madgwick:03}
Madgwick D.~S.,  et~al., 2003, \mnras, 344, 847

\bibitem[\protect\citeauthoryear{{Maraston}}{{Maraston}}{2005}]{Maraston:05}
{Maraston} C.,  2005, \mn@doi [\mnras] {10.1111/j.1365-2966.2005.09270.x},
  \href {https://ui.adsabs.harvard.edu/abs/2005MNRAS.362..799M} {362, 799}

\bibitem[\protect\citeauthoryear{{Maraston} \& {Str{\"o}mb{\"a}ck}}{{Maraston}
  \& {Str{\"o}mb{\"a}ck}}{2011}]{Maraston:11}
{Maraston} C.,  {Str{\"o}mb{\"a}ck} G.,  2011, \mn@doi [\mnras]
  {10.1111/j.1365-2966.2011.19738.x}, \href
  {https://ui.adsabs.harvard.edu/abs/2011MNRAS.418.2785M} {418, 2785}

\bibitem[\protect\citeauthoryear{{Melchior}, {Liang}, {Hahn}  \&
  {Goulding}}{{Melchior} et~al.}{2022}]{Melchior:22}
{Melchior} P.,  {Liang} Y.,  {Hahn} C.,   {Goulding} A.,  2022, \mn@doi [arXiv
  e-prints] {10.48550/arXiv.2211.07890}, \href
  {https://ui.adsabs.harvard.edu/abs/2022arXiv221107890M} {p. arXiv:2211.07890}

\bibitem[\protect\citeauthoryear{{Mo}, {Mao}  \& {White}}{{Mo}
  et~al.}{1998}]{MMW:98}
{Mo} H.~J.,  {Mao} S.,   {White} S. D.~M.,  1998, \mn@doi [\mnras]
  {10.1046/j.1365-8711.1998.01227.x}, \href
  {https://ui.adsabs.harvard.edu/abs/1998MNRAS.295..319M} {295, 319}

\bibitem[\protect\citeauthoryear{{Nersesian} et~al.,}{{Nersesian}
  et~al.}{2021}]{Nersesian:21}
{Nersesian} A.,  et~al., 2021, \mn@doi [\mnras] {10.1093/mnras/stab1984}, \href
  {https://ui.adsabs.harvard.edu/abs/2021MNRAS.506.3986N} {506, 3986}

\bibitem[\protect\citeauthoryear{{Nogueira-Cavalcante}, {Gon{\c{c}}alves},
  {Men{\'e}ndez-Delmestre}  \& {Sheth}}{{Nogueira-Cavalcante}
  et~al.}{2018}]{Nogueira:18}
{Nogueira-Cavalcante} J.~P.,  {Gon{\c{c}}alves} T.~S.,
  {Men{\'e}ndez-Delmestre} K.,   {Sheth} K.,  2018, \mn@doi [\mnras]
  {10.1093/mnras/stx2399}, \href
  {https://ui.adsabs.harvard.edu/abs/2018MNRAS.473.1346N} {473, 1346}

\bibitem[\protect\citeauthoryear{{Nolan}, {Raychaudhury}  \&
  {Kab{\'a}n}}{{Nolan} et~al.}{2007}]{Nolan:07}
{Nolan} L.~A.,  {Raychaudhury} S.,   {Kab{\'a}n} A.,  2007, \mn@doi [\mnras]
  {10.1111/j.1365-2966.2006.11326.x}, \href
  {https://ui.adsabs.harvard.edu/abs/2007MNRAS.375..381N} {375, 381}

\bibitem[\protect\citeauthoryear{{Pasquali}, {Gallazzi}, {Fontanot}, {van den
  Bosch}, {De Lucia}, {Mo}  \& {Yang}}{{Pasquali} et~al.}{2010}]{Pasquali:10}
{Pasquali} A.,  {Gallazzi} A.,  {Fontanot} F.,  {van den Bosch} F.~C.,  {De
  Lucia} G.,  {Mo} H.~J.,   {Yang} X.,  2010, \mn@doi [\mnras]
  {10.1111/j.1365-2966.2010.17074.x}, \href
  {https://ui.adsabs.harvard.edu/abs/2010MNRAS.407..937P} {407, 937}

\bibitem[\protect\citeauthoryear{{Peng}, {Lilly}, {Renzini}  \&
  {Carollo}}{{Peng} et~al.}{2012}]{Peng:12}
{Peng} Y.-j.,  {Lilly} S.~J.,  {Renzini} A.,   {Carollo} M.,  2012, \mn@doi
  [\apj] {10.1088/0004-637X/757/1/4}, \href
  {https://ui.adsabs.harvard.edu/abs/2012ApJ...757....4P} {757, 4}

\bibitem[\protect\citeauthoryear{{Portillo}, {Parejko}, {Vergara}  \&
  {Connolly}}{{Portillo} et~al.}{2020}]{Portillo:20}
{Portillo} S. K.~N.,  {Parejko} J.~K.,  {Vergara} J.~R.,   {Connolly} A.~J.,
  2020, \mn@doi [\aj] {10.3847/1538-3881/ab9644}, \href
  {https://ui.adsabs.harvard.edu/abs/2020AJ....160...45P} {160, 45}

\bibitem[\protect\citeauthoryear{{Renzini}}{{Renzini}}{2006}]{Renzini:06}
{Renzini} A.,  2006, \mn@doi [\araa] {10.1146/annurev.astro.44.051905.092450},
  \href {https://ui.adsabs.harvard.edu/abs/2006ARA&A..44..141R} {44, 141}

\bibitem[\protect\citeauthoryear{{Robotham}, {Bellstedt}, {Lagos}, {Thorne},
  {Davies}, {Driver}  \& {Bravo}}{{Robotham} et~al.}{2020}]{Robotham:20}
{Robotham} A.~S.~G.,  {Bellstedt} S.,  {Lagos} C. d.~P.,  {Thorne} J.~E.,
  {Davies} L.~J.,  {Driver} S.~P.,   {Bravo} M.,  2020, \mn@doi [\mnras]
  {10.1093/mnras/staa1116}, \href
  {https://ui.adsabs.harvard.edu/abs/2020MNRAS.495..905R} {495, 905}

\bibitem[\protect\citeauthoryear{{Rogers}, {Ferreras}, {Lahav}, {Bernardi},
  {Kaviraj}  \& {Yi}}{{Rogers} et~al.}{2007}]{Rogers:07}
{Rogers} B.,  {Ferreras} I.,  {Lahav} O.,  {Bernardi} M.,  {Kaviraj} S.,   {Yi}
  S.~K.,  2007, \mn@doi [\mnras] {10.1111/j.1365-2966.2007.12446.x}, \href
  {https://ui.adsabs.harvard.edu/abs/2007MNRAS.382..750R} {382, 750}

\bibitem[\protect\citeauthoryear{{Rogers}, {Ferreras}, {Peletier}  \&
  {Silk}}{{Rogers} et~al.}{2010a}]{BMC:10}
{Rogers} B.,  {Ferreras} I.,  {Peletier} R.,   {Silk} J.,  2010a, \mn@doi
  [\mnras] {10.1111/j.1365-2966.2009.15892.x}, \href
  {https://ui.adsabs.harvard.edu/abs/2010MNRAS.402..447R} {402, 447}

\bibitem[\protect\citeauthoryear{{Rogers}, {Ferreras}, {Pasquali}, {Bernardi},
  {Lahav}  \& {Kaviraj}}{{Rogers} et~al.}{2010b}]{Rogers:10}
{Rogers} B.,  {Ferreras} I.,  {Pasquali} A.,  {Bernardi} M.,  {Lahav} O.,
  {Kaviraj} S.,  2010b, \mn@doi [\mnras] {10.1111/j.1365-2966.2010.16436.x},
  \href {https://ui.adsabs.harvard.edu/abs/2010MNRAS.405..329R} {405, 329}

\bibitem[\protect\citeauthoryear{{Salim}}{{Salim}}{2014}]{Salim:14}
{Salim} S.,  2014, \mn@doi [Serbian Astronomical Journal]
  {10.2298/SAJ1489001S}, \href
  {https://ui.adsabs.harvard.edu/abs/2014SerAJ.189....1S} {189, 1}

\bibitem[\protect\citeauthoryear{{Salvador-Rusi{\~n}ol}, {Vazdekis}, {La
  Barbera}, {Beasley}, {Ferreras}, {Negri}  \& {Dalla
  Vecchia}}{{Salvador-Rusi{\~n}ol} et~al.}{2020}]{Salvador:20}
{Salvador-Rusi{\~n}ol} N.,  {Vazdekis} A.,  {La Barbera} F.,  {Beasley} M.~A.,
  {Ferreras} I.,  {Negri} A.,   {Dalla Vecchia} C.,  2020, \mn@doi [Nature
  Astronomy] {10.1038/s41550-019-0955-0}, \href
  {https://ui.adsabs.harvard.edu/abs/2020NatAs...4..252S} {4, 252}

\bibitem[\protect\citeauthoryear{{Schawinski}}{{Schawinski}}{2012}]{Schawinski:12}
{Schawinski} K.,  2012, \mn@doi [arXiv e-prints] {10.48550/arXiv.1206.2661},
  \href {https://ui.adsabs.harvard.edu/abs/2012arXiv1206.2661S} {p.
  arXiv:1206.2661}

\bibitem[\protect\citeauthoryear{{Schawinski}, {Thomas}, {Sarzi}, {Maraston},
  {Kaviraj}, {Joo}, {Yi}  \& {Silk}}{{Schawinski} et~al.}{2007}]{Schawinski:07}
{Schawinski} K.,  {Thomas} D.,  {Sarzi} M.,  {Maraston} C.,  {Kaviraj} S.,
  {Joo} S.-J.,  {Yi} S.~K.,   {Silk} J.,  2007, \mn@doi [\mnras]
  {10.1111/j.1365-2966.2007.12487.x}, \href
  {https://ui.adsabs.harvard.edu/abs/2007MNRAS.382.1415S} {382, 1415}

\bibitem[\protect\citeauthoryear{{Schawinski} et~al.,}{{Schawinski}
  et~al.}{2014}]{Schawinski:14}
{Schawinski} K.,  et~al., 2014, \mn@doi [\mnras] {10.1093/mnras/stu327}, \href
  {https://ui.adsabs.harvard.edu/abs/2014MNRAS.440..889S} {440, 889}

\bibitem[\protect\citeauthoryear{{Smee} et~al.}{{Smee} et~al.}{2013}]{Smee:13}
{Smee} S.~A.,  et~al., 2013, \mn@doi [\aj] {10.1088/0004-6256/146/2/32}, \href
  {https://ui.adsabs.harvard.edu/abs/2013AJ....146...32S} {146, 32}

\bibitem[\protect\citeauthoryear{{Strateva} et~al.}{{Strateva}
  et~al.}{2001}]{Strateva:01}
{Strateva} et~al., 2001, \mn@doi [\aj] {10.1086/323301}, \href
  {https://ui.adsabs.harvard.edu/abs/2001AJ....122.1861S} {122, 1861}

\bibitem[\protect\citeauthoryear{{Thomas}, {Maraston}, {Bender}  \& {Mendes de
  Oliveira}}{{Thomas} et~al.}{2005}]{Thomas:05}
{Thomas} D.,  {Maraston} C.,  {Bender} R.,   {Mendes de Oliveira} C.,  2005,
  \mn@doi [\apj] {10.1086/426932}, \href
  {https://ui.adsabs.harvard.edu/abs/2005ApJ...621..673T} {621, 673}

\bibitem[\protect\citeauthoryear{{Tojeiro}, {Heavens}, {Jimenez}  \&
  {Panter}}{{Tojeiro} et~al.}{2007}]{Vespa:07}
{Tojeiro} R.,  {Heavens} A.~F.,  {Jimenez} R.,   {Panter} B.,  2007, \mn@doi
  [\mnras] {10.1111/j.1365-2966.2007.12323.x}, \href
  {https://ui.adsabs.harvard.edu/abs/2007MNRAS.381.1252T} {381, 1252}

\bibitem[\protect\citeauthoryear{{Tous}, {Solanes}  \& {Perea}}{{Tous}
  et~al.}{2020}]{Tous:20}
{Tous} J.~L.,  {Solanes} J.~M.,   {Perea} J.~D.,  2020, \mn@doi [\mnras]
  {10.1093/mnras/staa1408}, \href
  {https://ui.adsabs.harvard.edu/abs/2020MNRAS.495.4135T} {495, 4135}

\bibitem[\protect\citeauthoryear{{Trager}, {Worthey}, {Faber}, {Burstein}  \&
  {Gonz{\'a}lez}}{{Trager} et~al.}{1998}]{Trager:98}
{Trager} S.~C.,  {Worthey} G.,  {Faber} S.~M.,  {Burstein} D.,   {Gonz{\'a}lez}
  J.~J.,  1998, \mn@doi [\apjs] {10.1086/313099}, \href
  {https://ui.adsabs.harvard.edu/abs/1998ApJS..116....1T} {116, 1}

\bibitem[\protect\citeauthoryear{{Trager}, {Faber}, {Worthey}  \&
  {Gonz{\'a}lez}}{{Trager} et~al.}{2000}]{Trager:00}
{Trager} S.~C.,  {Faber} S.~M.,  {Worthey} G.,   {Gonz{\'a}lez} J.~J.,  2000,
  \mn@doi [\aj] {10.1086/301442}, \href
  {https://ui.adsabs.harvard.edu/abs/2000AJ....120..165T} {120, 165}

\bibitem[\protect\citeauthoryear{{Vazdekis}, {S{\'a}nchez-Bl{\'a}zquez},
  {Falc{\'o}n-Barroso}, {Cenarro}, {Beasley}, {Cardiel}, {Gorgas}  \&
  {Peletier}}{{Vazdekis} et~al.}{2010}]{Vazdekis:10}
{Vazdekis} A.,  {S{\'a}nchez-Bl{\'a}zquez} P.,  {Falc{\'o}n-Barroso} J.,
  {Cenarro} A.~J.,  {Beasley} M.~A.,  {Cardiel} N.,  {Gorgas} J.,   {Peletier}
  R.~F.,  2010, \mn@doi [\mnras] {10.1111/j.1365-2966.2010.16407.x}, \href
  {https://ui.adsabs.harvard.edu/abs/2010MNRAS.404.1639V} {404, 1639}

\bibitem[\protect\citeauthoryear{{Vazdekis}, {Ricciardelli}, {Cenarro},
  {Rivero-Gonz{\'a}lez}, {D{\'\i}az-Garc{\'\i}a}  \&
  {Falc{\'o}n-Barroso}}{{Vazdekis} et~al.}{2012}]{MIUSCAT}
{Vazdekis} A.,  {Ricciardelli} E.,  {Cenarro} A.~J.,  {Rivero-Gonz{\'a}lez}
  J.~G.,  {D{\'\i}az-Garc{\'\i}a} L.~A.,   {Falc{\'o}n-Barroso} J.,  2012,
  \mn@doi [\mnras] {10.1111/j.1365-2966.2012.21179.x}, \href
  {https://ui.adsabs.harvard.edu/abs/2012MNRAS.424..157V} {424, 157}

\bibitem[\protect\citeauthoryear{{Vazdekis}, {Koleva}, {Ricciardelli},
  {R{\"o}ck}  \& {Falc{\'o}n-Barroso}}{{Vazdekis} et~al.}{2016}]{AV:16}
{Vazdekis} A.,  {Koleva} M.,  {Ricciardelli} E.,  {R{\"o}ck} B.,
  {Falc{\'o}n-Barroso} J.,  2016, \mn@doi [\mnras] {10.1093/mnras/stw2231},
  \href {https://ui.adsabs.harvard.edu/abs/2016MNRAS.463.3409V} {463, 3409}

\bibitem[\protect\citeauthoryear{{Walcher}, {Groves}, {Budav{\'a}ri}  \&
  {Dale}}{{Walcher} et~al.}{2011}]{Walcher:11}
{Walcher} J.,  {Groves} B.,  {Budav{\'a}ri} T.,   {Dale} D.,  2011, \mn@doi
  [\apss] {10.1007/s10509-010-0458-z}, \href
  {https://ui.adsabs.harvard.edu/abs/2011Ap&SS.331....1W} {331, 1}

\bibitem[\protect\citeauthoryear{{Wild} \& {Hewett}}{{Wild} \&
  {Hewett}}{2005}]{WH:05}
{Wild} V.,  {Hewett} P.~C.,  2005, \mn@doi [\mnras]
  {10.1111/j.1365-2966.2005.08844.x}, \href
  {https://ui.adsabs.harvard.edu/abs/2005MNRAS.358.1083W} {358, 1083}

\bibitem[\protect\citeauthoryear{{Wild} et~al.,}{{Wild} et~al.}{2014}]{Wild:14}
{Wild} V.,  et~al., 2014, \mn@doi [\mnras] {10.1093/mnras/stu212}, \href
  {https://ui.adsabs.harvard.edu/abs/2014MNRAS.440.1880W} {440, 1880}

\bibitem[\protect\citeauthoryear{{Wilkinson}, {Maraston}, {Goddard}, {Thomas}
  \& {Parikh}}{{Wilkinson} et~al.}{2017}]{Firefly:17}
{Wilkinson} D.~M.,  {Maraston} C.,  {Goddard} D.,  {Thomas} D.,   {Parikh} T.,
  2017, \mn@doi [\mnras] {10.1093/mnras/stx2215}, \href
  {https://ui.adsabs.harvard.edu/abs/2017MNRAS.472.4297W} {472, 4297}

\bibitem[\protect\citeauthoryear{{Worthey}}{{Worthey}}{1994}]{W:94}
{Worthey} G.,  1994, \mn@doi [\apjs] {10.1086/192096}, \href
  {https://ui.adsabs.harvard.edu/abs/1994ApJS...95..107W} {95, 107}

\bibitem[\protect\citeauthoryear{{Worthey} \& {Ottaviani}}{{Worthey} \&
  {Ottaviani}}{1997}]{WO:97}
{Worthey} G.,  {Ottaviani} D.~L.,  1997, \mn@doi [\apjs] {10.1086/313021},
  \href {https://ui.adsabs.harvard.edu/abs/1997ApJS..111..377W} {111, 377}

\bibitem[\protect\citeauthoryear{{Yip} et~al.,}{{Yip} et~al.}{2004}]{Yip:04}
{Yip} C.~W.,  et~al., 2004, \mn@doi [\aj] {10.1086/425626}, \href
  {https://ui.adsabs.harvard.edu/abs/2004AJ....128.2603Y} {128, 2603}

\bibitem[\protect\citeauthoryear{{York} et~al.}{{York} et~al.}{2000}]{SDSS}
{York} D.~G.,  et~al., 2000, \mn@doi [\aj] {10.1086/301513}, \href
  {https://ui.adsabs.harvard.edu/abs/2000AJ....120.1579Y} {120, 1579}

\bibitem[\protect\citeauthoryear{{de La Rosa}, {La Barbera}, {Ferreras}  \& {de
  Carvalho}}{{de La Rosa} et~al.}{2011}]{IGDR:11}
{de La Rosa} I.~G.,  {La Barbera} F.,  {Ferreras} I.,   {de Carvalho} R.~R.,
  2011, \mn@doi [\mnras] {10.1111/j.1745-3933.2011.01146.x}, \href
  {https://ui.adsabs.harvard.edu/abs/2011MNRAS.418L..74D} {418, L74}

\bibitem[\protect\citeauthoryear{{van Dokkum} \& {Conroy}}{{van Dokkum} \&
  {Conroy}}{2010}]{vDC:10}
{van Dokkum} P.~G.,  {Conroy} C.,  2010, \mn@doi [\nat] {10.1038/nature09578},
  \href {https://ui.adsabs.harvard.edu/abs/2010Natur.468..940V} {468, 940}

\makeatother
\end{thebibliography}

\begin{appendix}

\section{Assessing the potential effect of a sky residual}
\label{Sec:ApSky}
\begin{figure*}
\includegraphics[width=85mm]{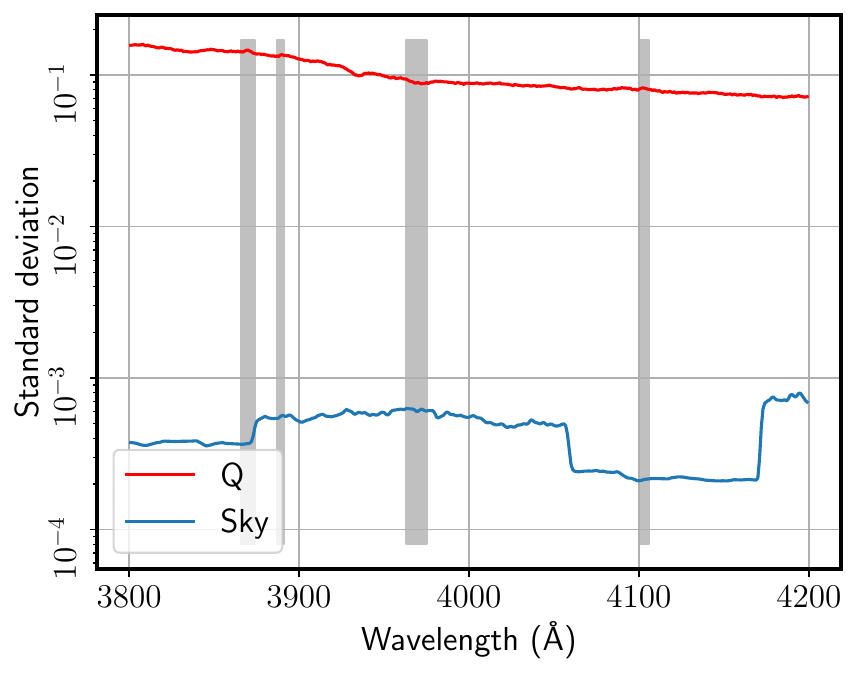}
\includegraphics[width=85mm]{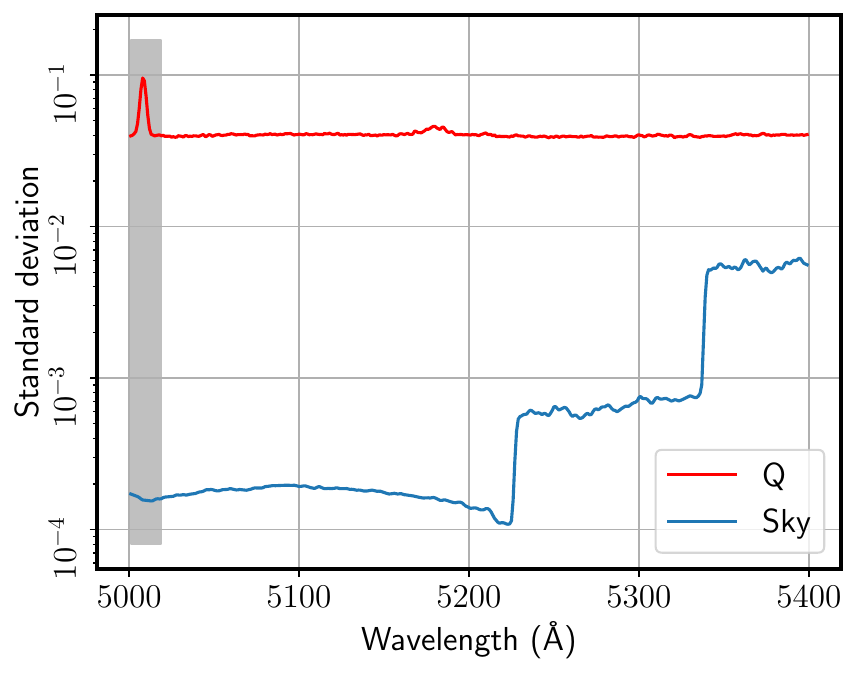}
\caption{Comparison of the standard deviation in the quiescent
  galaxies (red lines) with that of the sky spectrum (blue lines) of
  the same galaxies, combined according to the redshift distribution,
  i.e. the sky spectra are brought to the rest-frame for each
  galaxy. The red lines are thus equivalent to those shown in the
  bottom panels of Fig.~\ref{fig:SigPlot}. The left (right) panels
  correspond to the blue (red) spectral intervals adopted in this
  work. Note this is an unrealistic worst case scenario, as it
  corresponds to spectra without any sky subtraction.}
\label{fig:Sky}
\end{figure*}

The study presented here is based on the science-ready spectra
from the Legacy-SDSS survey. These data have been shown in
many papers to be of the highest quality for detailed
analyses of large samples, including work on stacked data
\citep[e.g.,][]{Gallazzi:05,Thomas:05,Rogers:07,FLB:13,FLB:14,SDSSCPs}.
However, since PCA is overly sensitive to any
potential systematic, in this appendix we focus on the issue
of sky subtraction, giving a quantitative upper bound on the
contribution of sky residuals to the variance of the set. In
Sec.~\ref{Sec:Dec}, we argued that, due to the wide distribution
of redshifts of our sample, and the narrowness of the airglow lines,
the net effect of residuals from sky subtraction are expected to
be small, and not drive the signal in the highest principal components.
We test this issue by taking a typical sky spectrum from the
SDSS database \citep[see, e.g.,][]{Hart:19}, producing sky spectra in the rest-frame of the
galaxies according to the redshift distribution of the sample.
For simplicity, we restrict the analysis to quiescent galaxies, although the
result is very similar for the other subsets. Fig.~\ref{fig:Sky}
compares the standard deviation of the sample -- in red, identical to the red
curves in the bottom panels of Fig.~\ref{fig:SigPlot} -- with that of the
distribution of sky spectra (blue),  noting that we also normalize the airglow 
spectra using the galaxy flux in the 6000-6500\AA\ for consistency.
The left (right) panels correspond to the blue (red) spectral
intervals explored in this work. The contribution from the sky is much lower,
representing a fraction of the galaxy variance between 1/300 for the blue
interval and 1/200 for the red interval.  We emphasize that this
comparison just shows the net maximum variance contributed by the sky
spectrum to the analysis if it were not subtracted from the data. A
typical residual in the sky-subtracted spectrum below the $\sim$5 percent
level will produce a contribution to the standard deviation $\sim$20 times
smaller.  To confirm this point, we note that the eigenvectors shown
in Fig.~\ref{fig:eigenS} do
not inherit the spectral structure from airglow, as one would expect
if the sky were to contribute significantly to the analysis. In
particular this is more obvious in the red interval, where the airglow
would introduce a large amount of variance in the redder part of the
spectrum. While still in preparation, we also note that comparisons
with synthetic spectra from hydrodynamical simulations of galaxy
formation -- that do not have any sky contaminantion added in the
process, produce a distribution of spectra that is statistically very
similar to the results presented here (Sharbaf et al., in
preparation).

\section{Spectral fitting for PCA in the red interval}
\label{Sec:ApSpecRed}

\begin{figure*}
\includegraphics[width=.05\linewidth]{SSPFits/Row0.png}
\includegraphics[width=.3\linewidth]{SSPFits/Col1.png}
\includegraphics[width=.3\linewidth]{SSPFits/Col2.png}
\includegraphics[width=.3\linewidth]{SSPFits/Col3.png}
\includegraphics[width=.05\linewidth]{SSPFits/Row1.png}
\includegraphics[width=.3\linewidth]{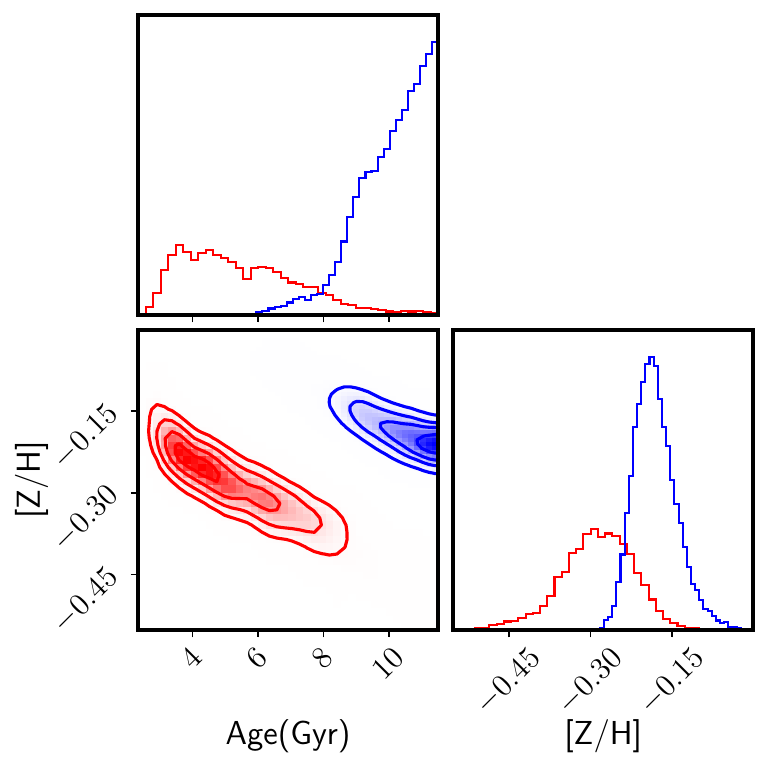}
\includegraphics[width=.3\linewidth]{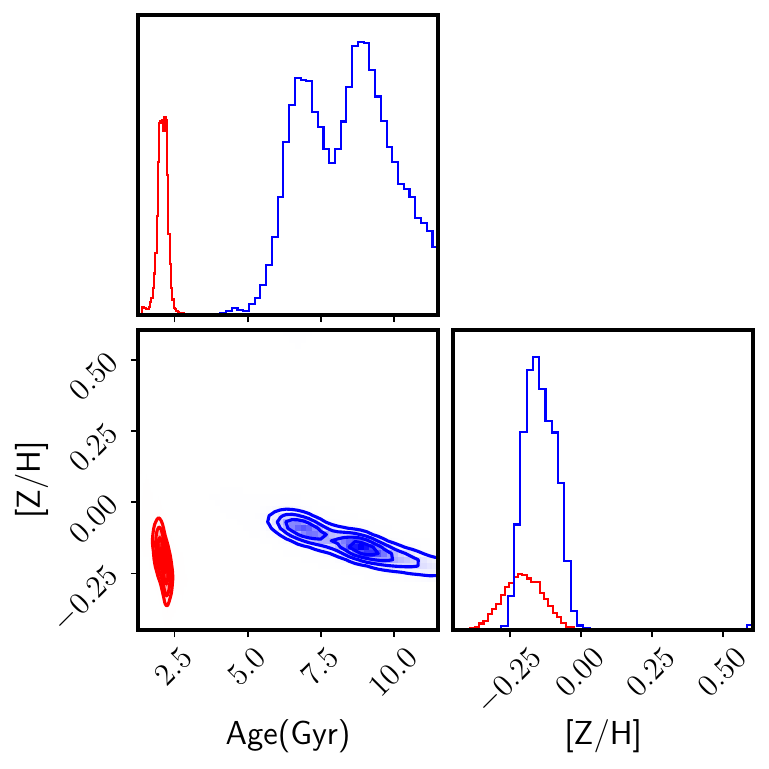}
\includegraphics[width=.3\linewidth]{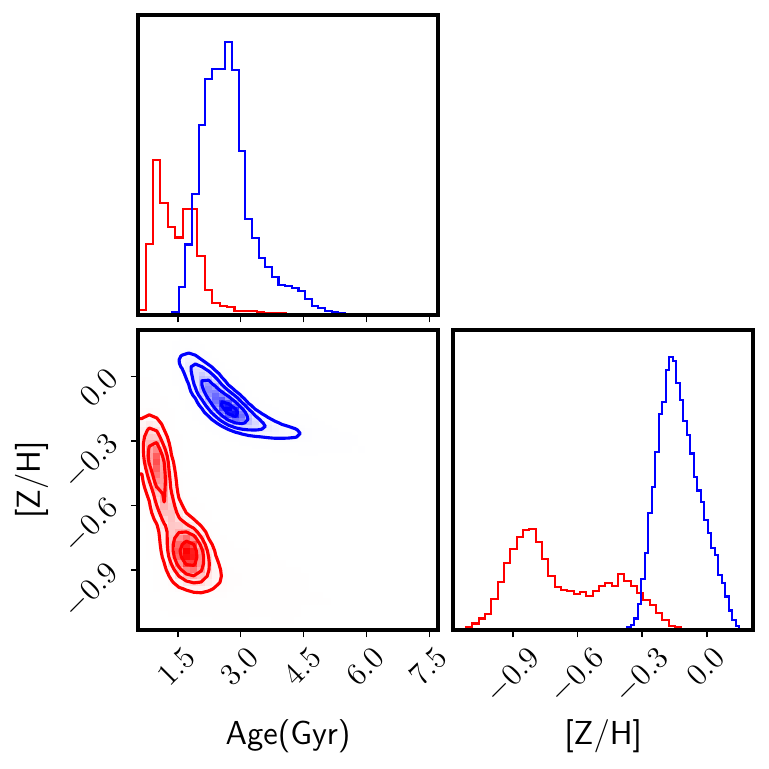}
\includegraphics[width=.05\linewidth]{SSPFits/Row2.png}
\includegraphics[width=.3\linewidth]{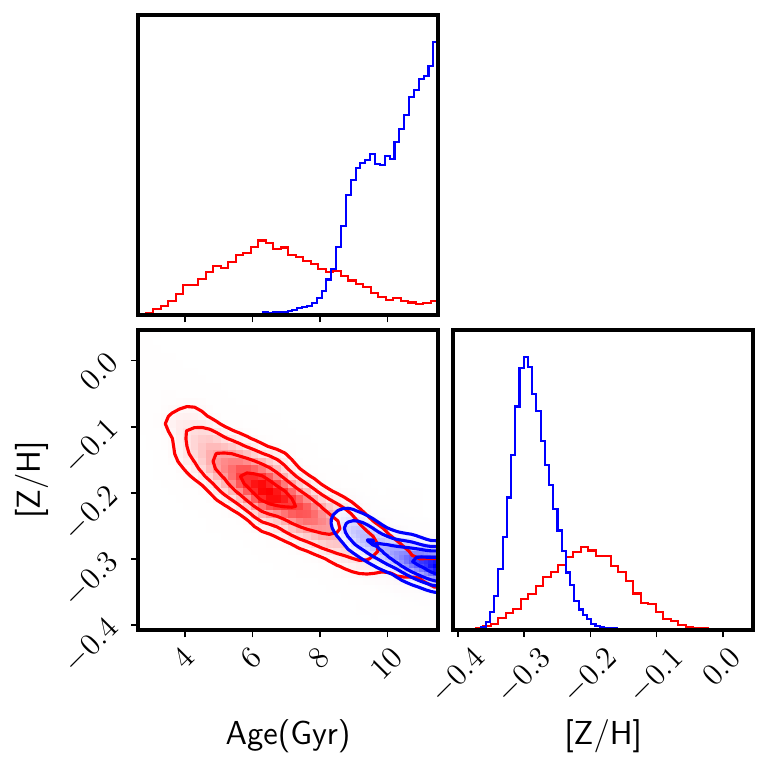}
\includegraphics[width=.3\linewidth]{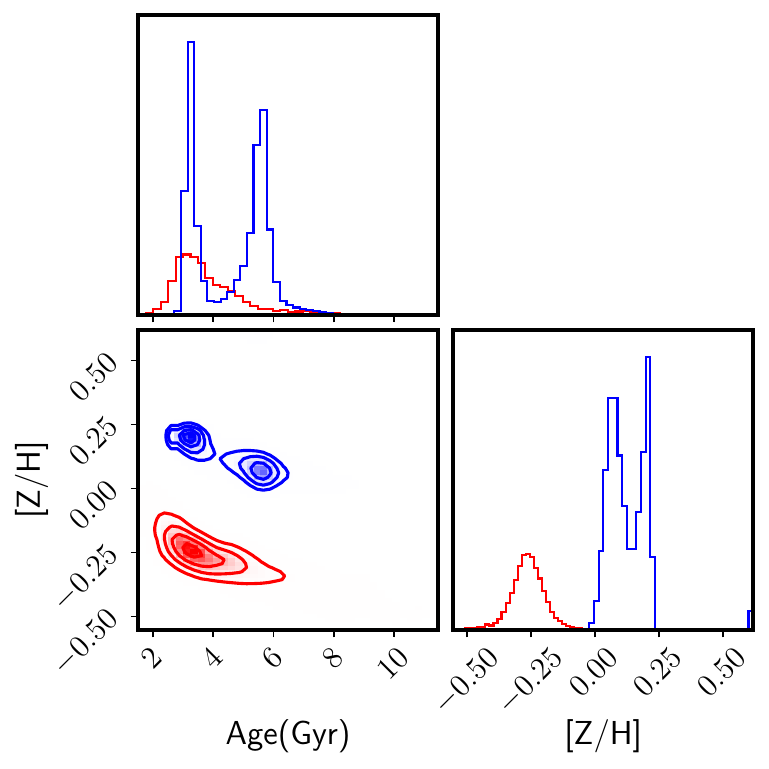}
\includegraphics[width=.3\linewidth]{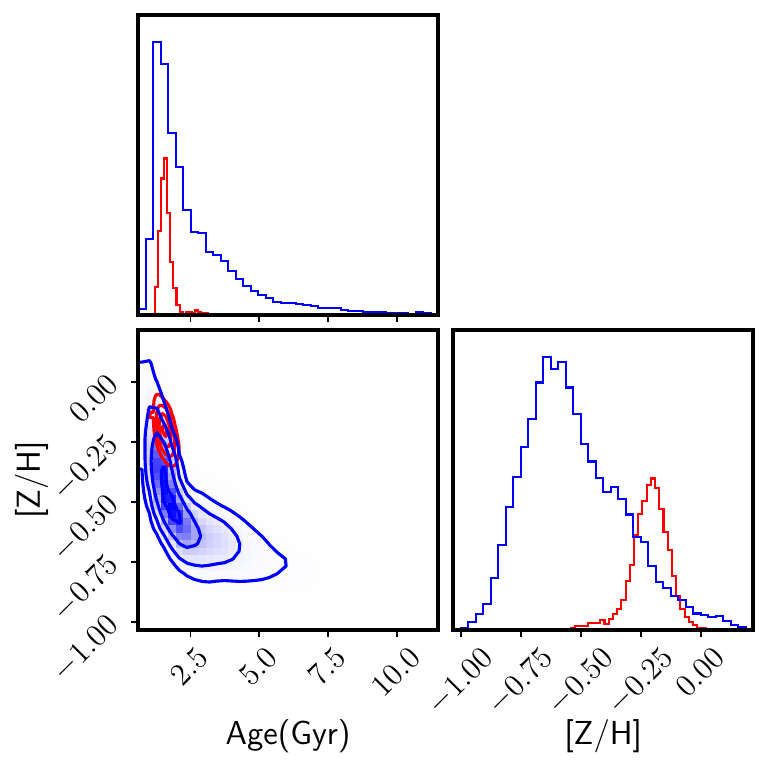}
\includegraphics[width=.05\linewidth]{SSPFits/Row3.png}
\includegraphics[width=.3\linewidth]{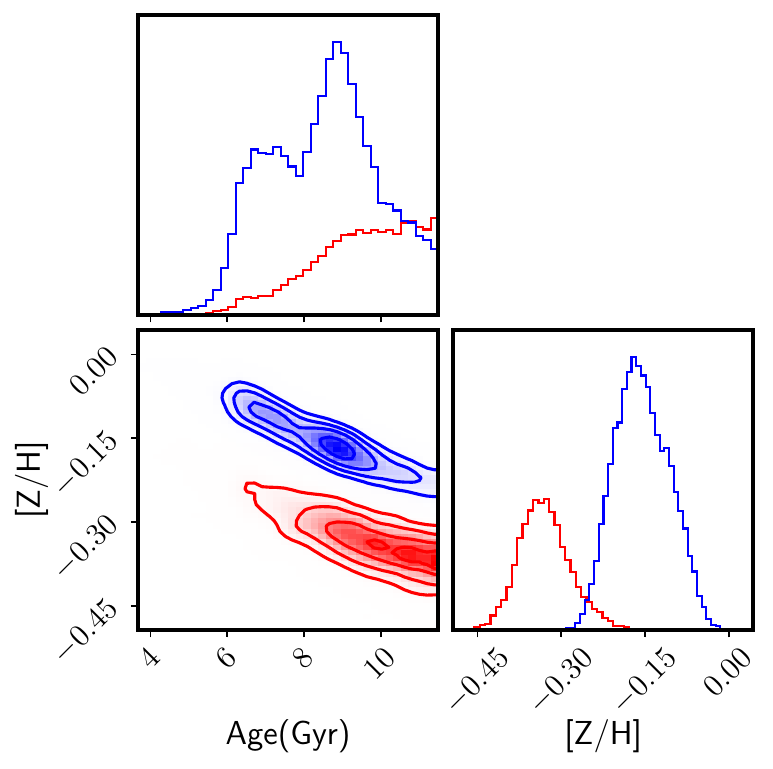}
\includegraphics[width=.3\linewidth]{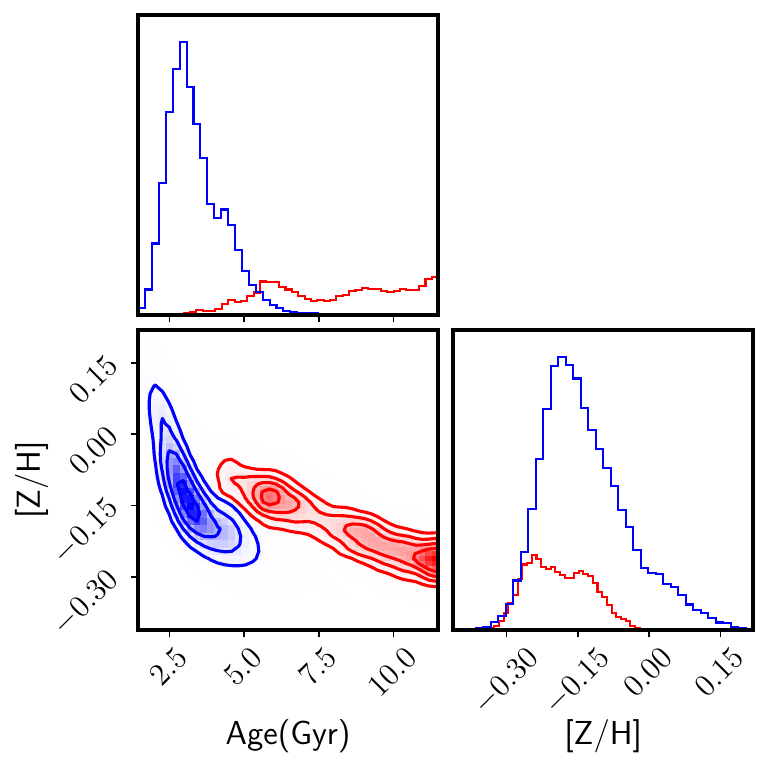}
\includegraphics[width=.3\linewidth]{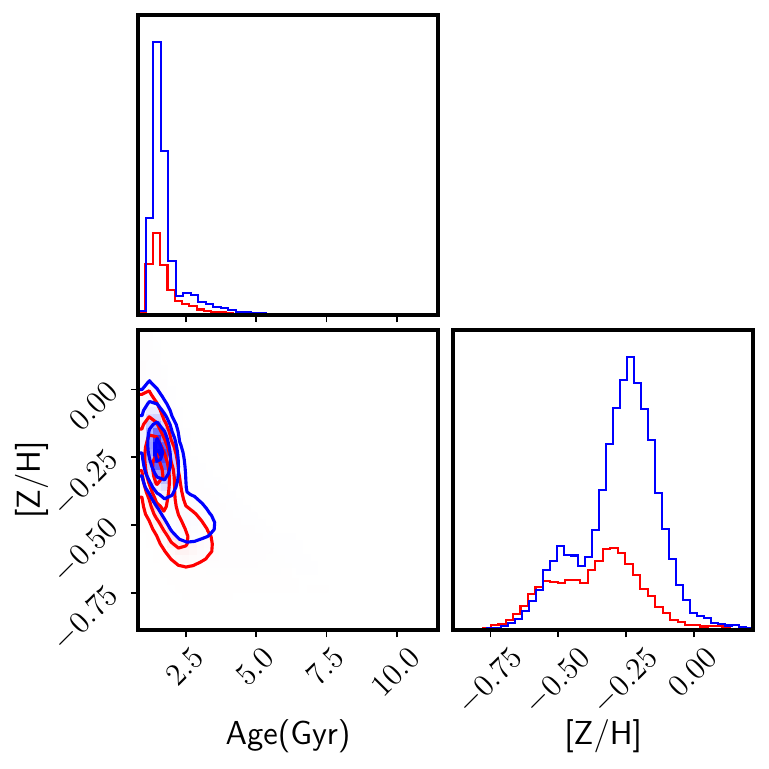}
\caption{Equivalent of Fig.~\ref{fig:MCMC} when adopting the
  red interval (5000-5400\AA).}
\label{fig:MCMCRed}
\end{figure*}

Concerning the result of the spectral fitting of stacked data
according to extreme values of the principal component projections,
we show in Fig.~\ref{fig:MCMCRed} the equivalent of Fig.~\ref{fig:MCMC}
for the analysis perfomed in the red interval (5000-5400\AA), using the
same fitting models and parameter ranges.

\end{appendix}

\bsp	
\label{lastpage}
\end{document}